\documentclass{jfm}

\usepackage{graphicx}
\usepackage{bm}
\usepackage{epstopdf,epsfig}
\usepackage{newtxtext}
\usepackage{newtxmath}
\usepackage{natbib}
\usepackage{hyperref}
\hypersetup{
    colorlinks = true,
    urlcolor   = blue,
    citecolor  = black,
}

\newcommand{\RomanNumeralCaps}[1]

\usepackage{algorithm}
\usepackage{algpseudocode}

\newcommand{\bsq}{\boldsymbol{q}}

\newcommand{\bsg}{\boldsymbol{g}}

\newcommand{\bsqt}{\hat{\boldsymbol{q}}}
\newcommand{\bsf}{\boldsymbol{f}}
\newcommand{\bmu}{\boldsymbol{\mu}}
\newcommand{\bsft}{\hat{\boldsymbol{f}}}
\newcommand{\bsx}{\boldsymbol{x}}

\newcommand{\bsU}{\boldsymbol{U}}

\newcommand{\bsu}{\boldsymbol{u}}
\newcommand{\bsw}{\boldsymbol{w}}

\newcommand{\bomega}{\boldsymbol{\omega}}

\newcommand{\pp}{\partial}
\DeclareMathOperator*{\argmax}{arg\max}
\DeclareMathOperator*{\argmin}{arg\min}

\graphicspath{{Figures/}}

\title{Sparse space-time resolvent analysis for statistically-stationary and time-varying flows
}

\author{Barbara Lopez-Doriga\aff{1}
  \corresp{\email{blopezdorigacostales@hawk.iit.edu}},
  Eric Ballouz\aff{2},
  H. Jane Bae\aff{3},
 \and Scott T. M. Dawson\aff{1}}

\affiliation{\aff{1}Mechanical, Materials, and Aerospace Engineering Department, Illinois Institute of Technology, Chicago, IL 60616, USA
\aff{2}Mechanical and Civil Engineering, California Institute of Technology, Pasadena, CA, 91125, USA
\aff{3}Graduate Aerospace Laboratory, California Institute of Technology, Pasadena, CA, 91125, USA}

\begin{document}
\maketitle

\begin{abstract}
Resolvent analysis provides a framework to predict coherent spatio-temporal structures of largest linear energy amplification, through a singular value decomposition (SVD) of the resolvent operator, obtained by linearizing the Navier--Stokes equations about a known turbulent mean velocity profile. Resolvent analysis utilizes a Fourier decomposition in time, which has thus-far limited its application to statistically-stationary or time-periodic flows. This work develops a variant of resolvent analysis applicable to time-evolving flows, and proposes a variant that identifies spatio-temporally sparse structures, applicable to either stationary or time-varying mean velocity profiles. Spatio-temporal  resolvent analysis is formulated through the incorporation of the temporal dimension to the numerical domain via a discrete time-differentiation operator. Sparsity (which manifests in localisation) is achieved through the addition of an $l_1$-norm penalisation term to the optimisation associated with the SVD. This modified optimization problem can be formulated as a nonlinear eigenproblem,  and solved via an inverse power method. We first showcase the implementation of the sparse analysis on statistically-stationary turbulent channel flow, and demonstrate that the sparse variant can identify aspects of the physics not directly evident from standard resolvent analysis. This is followed by applying the sparse space-time formulation on systems that are time-varying: a time-periodic turbulent Stokes boundary layer, and then a turbulent channel flow with a sudden imposition of a lateral pressure gradient, with the original streamwise pressure gradient unchanged. We present results demonstrating how the sparsity-promoting variant can either change the quantitative structure of the leading space-time modes to increase their sparsity, or identify entirely different linear amplification mechanisms compared to non-sparse resolvent analysis.
\end{abstract}

\begin{keywords}
\end{keywords}

\section{Introduction}
Despite the highly nonlinear nature of turbulent fluid flows, linearised analyses of the governing Navier--Stokes equations have proven to be effective at capturing several pertinent properties of such systems. Resolvent analysis, first applied as a model for turbulent pipe flow \citep{mckeon2010resolvent}, has been informative in other problems that involve wall-bounded turbulence, spatio-temporal flow statistics \citep{towne2020resolvent}, and the identification of coherent structures in turbulent jets \citep{lesshafft2019resolvent,pickering2021resolvent}, airfoils \citep{yeh2019resolvent} supersonic boundary layers \citep{bae2020resolvent,bae2020cooled} and turbulent rectangular duct flows \citep{ldoriga2022duct}. Combining a set of triadically consistent resolvent modes has also been used for the representation of hairpin structures in \citet{sharma2013resolvent}, and more generally for the reconstruction of phenomena observed in wall-bounded turbulence \citep{mckeon2017engine}. This framework has further been applied for the estimation of flow states \citep{gomez2016reduced,beneddine2017unsteady,symon2019mean,illingworth2018estimating}, the prediction of coherent structures \citep{abreu2020spectral,tissot2021stochastic} and statistical quantities and scalings \citep{hwang2010linear,zare2017colour,towne2020resolvent}, designing  control strategies for drag reduction \citep{luhar2014opposition,toedtli2019predicting}, and modelling the effect of complex surfaces \citep{luhar2015compliant,chavarin2019riblet}. The broad applicability of such linearised analysis relies on (and can be seen as evidence to infer) the importance of linear amplification mechanisms in the generation and evolution of empirically-observed coherent structures within turbulent flows, such as near-wall streaks \citep{kline1967structure}, hairpin vortices \citep{theodorsen1952mechanisms,head1981new}, superstructures \citep{kim1999very}, and a range of other coherent features described in \citet{jimenez2018coherent}.
 
The linearised analyses discussed thus far assume that the linear system under investigation is time-invariant, as is the case when the underlying flow is statistically stationary. The recently-developed harmonic resolvent analysis \citep{harmonic2020padovan,padovan2022analysis} enables the resolvent framework to extend to statistically time-periodic flows, enabling cross-frequency analysis capturing triadic interactions between a time-periodic base flow and fluctuations about this mean state at other frequencies. In the context of flow control over a periodically plunging cylinder, \citet{lin2023plungingcylinder} utilizes a  Lyapunouv–-Floquet transformation to map the corresponding linear time-periodic system to a time-invariant equivalent, enabling the application of standard resolvent analysis methods. Other noteworthy contributions to the study of time-varying linear systems include the linear stability analyses compiled in \cite{stabilityStokes2002,oscillatorychannel2006,oscillatorycylinder2007}, for flat and high-frequency oscillatory Stokes boundary layers, as well as an oscillating cylinder.

Whether considering a statistically-stationary or a time-periodic mean state, the methods discussed thus far consider a Fourier decomposition in time. Typically, this involves identifying the forcing (input) and response (output) structures corresponding to largest energy amplification by the linearized system (represented by the resolvent operator). 
While this decomposition arises naturally for such methods, it can potentially obscure the intermittent nature of velocity fluctuations present in turbulent flows. Alternative linear analyses methods can be similarly restrictive, with asymptotic stability analysis also identifying eigenmodes each associated with a single (possibly complex) frequency. Conversely, transient growth analysis \citep{boberg1988onset,butler1992optimal,reddy1993energy,Schmid:2007} considers the unforced response to a specific initial condition, corresponding to maximal energy growth over a specified time horizon. This again is unrealistic for systems subject to continuous perturbations \citep{jovanovic2005componentwise}, though such analysis has been used in turbulent flows, such as to predict the emergence of near-wall streamwise streak \citep{delAlamo2006linearAmp} and vortices \citep{schoppa2002coherent} in wall-bounded turbulence. 
To overcome these limitations, here we introduce a space-time formulation of the resolvent operator that is firstly applicable to non-statistically-stationary systems with an arbitrarily time-varying mean profile, and secondly allows for the identification of optimal input and output trajectories that can have arbitrary time dependence. This builds upon preliminary work first reported in \citet{ldoriga2023sparse}. While not explored here, related work also considers explicitly replacing the Fourier transform used in standard resolvent analysis with a wavelet transform \citep{ballouz2023wavelet,ballouz2024wavelet}.

This generalization of operator-based decompositions to enable non-Fourier temporal modes is somewhat analogous to efforts to similarly generalize data-driven proper orthogonal decomposition POD methodology to identify intermittent behavior in turbulent flows, such as the conditional POD formulated in \citet{schmidt2019conditional}, and time-windowed space-time POD described in \citet{frame2022space}. Note that spectral POD \citep{towne2018spectral} has also been recently generalized for time-periodic systems using cyclostationary analysis  \citep{heidt2023spectral}.

Methods to identify nonmodal linear energy amplification such as resolvent or transient growth analysis involve computing the leading singular values and vectors of an appropriately-defined linear operator. 
The singular value decomposition, by design, is defined as an optimisation problem that involves an $l_2$-energy norm. In the context of resolvent analysis, this optimization problem relates to the energy ratio between input and output flow states, and naturally yields spatio-temporal structures that are Fourier modes in time. 
Here, we consider modifications to the standard optimization problem that yield alternative temporal functions, which are inclined to be localised in time. This is achieved by incorporating an $l_1$ norm term into the optimisation problem. 
 The use of $l_1$ norms to promote localisation and/or sparsity has origins in compressive sensing \citep{candes2008introduction}.

In the context of fluid mechanics, sparsity-promoting methods have been utilised for developing reduced-complexity models across a number of contexts. These include the identification of sparse nonlinear reduced-order models \citep{brunton2016sindy,loiseau2018constrained,rubini2020l1}, the selection of a sparse set of active dynamic modes \citep{jovanovic2014dmdsp}, and in reconstruction of temporal spectral content from data that is under-resolved in time \citep{tu2014compressed}. Recently, sparsity promoting methods have also been incorporated  in the resolvent analysis framework in \citet{skene2022sparseForcing}, where they are used to identify spatially-localized forcing modes, which can be more directly useful for actuator placement in flow control applications. In \citet{skene2022sparseForcing}, a Riemannian optimization process is used to solve an $l_1$-based optimisation problem, following a similar approach used by \citet{foures2013localization}  to identify spatially-localised structures in transient growth analysis. The present work is similarly motivated, though we focus here on achieving localisation in time as well as space. We also use a different formulation of the optimisation problem, which allows for a balance between $l_1$ and $l_2$ norm contributions.

The structure of the paper is as follows. A discussion of the fundamentals of pseudospectral analysis and the wall-normal derivation of the governing equations, along with the space-time form of the resolvent operator, and a description of the algorithm that promotes sparsity on the resolvent modes are presented in \S\ref{sec:method}. The main results of our investigation are discussed in \S\ref{sec:results}: the sparse formulation of the standard resolvent operator is applied in the streamwise and spanwise directions of a turbulent channel flow in \S\ref{sec:resultsChannel1d}; and the space-time and sparse space-time formulations of the resolvent operator are applied on a turbulent channel flow in \S\ref{sec:resultsChannel2d}; a turbulent Stokes boundary layer in \S\ref{sec:resultsStokesBL}; and a channel flow with sudden lateral pressure gradient in \S\ref{sec:results3Dchannel}. Finally, we discuss the main findings and future prospects of our investigation in \S\ref{sec:conclusions}.

\section{Methodology}
\label{sec:method}
This section begins with a brief overview of the fundamentals of pseudospectral analysis of linear operators in \S\ref{sec:pseudo}. This is followed by a derivation of the resolvent formulation of the incompressible Navier--Stokes equations in wall-normal velocity and vorticity variables in \S\ref{sec:resolvent}, assuming homogeneity in both the spatial and temporal dimensions. This is followed by the development of a space-time resolvent operator where homogeneity is not assumed in the temporal dimension in \S\ref{sec:spaceTimeResolvent}, also in wall-normal velocity and vorticity variables. Following this, \S\ref{sec:sparseResolvent} introduces a formulation of resolvent analysis that promotes sparsity on the optimal resolvent modes. 

\subsection{Pseudospectral analysis of linear operators}
\label{sec:pseudo}
Let us consider a dynamical system governed by 
\begin{equation}
\label{eq:dynsys}
    \dot{\bsq} (\bsx,t)+ \mathcal{L} \bsq(\bsx,t) = \bsf(\bsx,t),
\end{equation}
where $\bsq$ denotes the state of the system with respect to a reference state $\bsq_0$, $\mathcal{L}$ is a linear operator and $\bsf$ represents an exogenous input or forcing. The space and time dimensions are denoted by $\bsx$ and $t$, respectively. Assuming that the system is homogeneous in the temporal dimension, we propose solutions of the form \citep{Schmid_Henningson} $\bsq(\bsx,t)=\bsqt(\bsx)\exp{(-i\omega t)}$ with $\omega \in \mathbb{C}$, and substituting in \eqref{eq:dynsys} gives
\begin{equation}
\label{eq:resolventInv}
    (-i\omega \mathsfbi{I}  + \mathcal{L})\bsqt(\bsx) = \bsft(\bsx).
\end{equation}
In the case the forcing term is nonzero, the elements can be rearranged so that the governing equation represents the following system
\begin{equation}
\label{eq:resolventDynSys}
    \bsqt(\bsx) = (-i\omega \mathsfbi{I}  + \mathcal{L})^{-1
    }\bsft(\bsx):=\mathcal{H}_\omega \bsft(\bsx).
\end{equation}
We refer to $\mathcal{H}_\omega$ as the resolvent operator. Note that the subscript $\omega$ is retained to highlight the dependence on the temporal frequency. The original dynamical system in \eqref{eq:dynsys} has been recast as a linear mapping between a forcing $\bsft$ and the  state $\bsqt$.  

According to \eqref{eq:resolventDynSys}, the properties of the state $\bsqt$ will be affected by both the nature of the forcing $\bsft$ and the properties of the resolvent $\mathcal{H}_\omega$. In this work, we focus in particular on the pairs of forcing-response that produce the largest amplification through the action of $\mathcal{H}_\omega$. That is, a forcing of small magnitude yields a response of large magnitude. Such structures can be identified via a singular value decomposition (SVD) of the resolvent operator $\mathcal{H}_\omega$ as follows
\begin{equation}
    \label{eq:svd_sum}
    \mathcal{H}_\omega = \sum_{j=1}^N \boldsymbol{\psi}_j \sigma_j \boldsymbol{\phi}_j^*,
\end{equation}
where $\sigma_j \geq \sigma_{j+1} \geq 0$ for all $j$, and $(\cdot^*)$ denotes the adjoint. Notice that here the resolvent operator will take the form of a discretised operator, therefore the summation in \eqref{eq:svd_sum} is truncated to $N$ terms. 

In particular, we seek $\hat \bsf = \boldsymbol{\phi}_1$ that maximises the largest singular value $\sigma_1$, where
\begin{equation}
\label{eq:sigopt}
    \sigma_1 = \max_{\boldsymbol{\phi}} \frac{\|\mathcal{H}_{\omega} \boldsymbol{\phi}\|_2}{\|\boldsymbol{\phi}\|_2}=\min_{\boldsymbol{\psi}}\frac{\|\boldsymbol{\psi}\|_2}{\|\mathcal{H}_{\omega}^*\boldsymbol{\psi}\|_2},
\end{equation}
where the $l_2$ norm is taken over the spatial domain $\Omega_{\bsx}$, so that, for example,
\begin{equation}
\label{eq:space2norm}
    \|\bm\psi(\bsx) \|_2 = \left(\int_{\bsx\in \Omega_{\bsx}} \left|\bm\psi(\bsx)\right|^2 \text{d}\bsx\right)^{1/2}.
\end{equation}
Alternatively, we can write this optimisation problem in terms of the leading forcing mode $\boldsymbol{\phi}_1$ as
\begin{equation}
\label{eq:phiopt}
    \boldsymbol{\phi}_1 = \argmax_{\boldsymbol{\phi}} \frac{\|\mathcal{H}_{\omega} \boldsymbol{\phi}\|_2}{\|\boldsymbol{\phi}\|_2},
\end{equation} 
or the leading response mode $\boldsymbol{\psi}_1$
\begin{equation}
\label{eq:psiopt}
\psi_1 = \argmin_\psi \frac{\|\psi\|_2}{\|\mathcal{H}_{\omega}^*\psi\|_2}.
\end{equation}

\subsection{Resolvent formulation of the mean-linearised incompressible Navier--Stokes equations}
\label{sec:resolvent}
The incompressible Navier--Stokes equations enforce conservation of momentum and mass, respectively, and are written in a Cartesian coordinate system as follows 
\begin{equation}
\label{eq:momentum}
    \pp_t \bsu = -\bsu \cdot \nabla\bsu-\nabla p+\frac{1}{\Rey}\Delta\bsu,
\end{equation}
\begin{equation}
\label{eq:cont}
    \nabla\cdot\bsu=0.
\end{equation}
Here, the instantaneous velocity field has three components: $\bsu=[u(\bsx,t),v(\bsx,t),$ $w(\bsx,t)]^T$ with $\bsx=[x,y,z]^T$ and $p=p(\bsx,t)$ represents the instantaneous pressure field. In this reference frame, $x$ and $z$ correspond to the streamwise and spanwise directions, respectively, and are nominally considered to be infinite in extent. The other variable, $y$, represents the wall-normal dimension. Here $\partial_t$ denotes a time (partial) derivative, the spatial gradient operator is given by $\nabla = [\partial_x,\partial_y,\partial_z]^T$, and the Laplacian operator is defined as $\Delta = \nabla \cdot \nabla$.

We can write a given instantaneous velocity state $\bsu$ as the sum of the temporal mean $\bsU$ and a fluctuating component $\bsu'$, such that
\begin{equation}
\label{eq:instantaneous}
    \bsu(\bsx,t) = \bsU(\bsx)+\bsu'(\bsx,t).
\end{equation}
 Applying this decomposition in \eqref{eq:momentum}--\eqref{eq:cont} and subtracting the temporal average gives the governing equations used in this work,
\begin{equation}
\label{eq:momentum3}
    \pp_t \bsu' +\bsU\cdot \nabla\bsu' +\bsu'\cdot \nabla\bsU + \nabla p'-\frac{1}{\Rey} \Delta\bsu' = -\bsu'\cdot\nabla\bsu'-\overline{\bsu'\cdot\nabla\bsu'}=\bsf',
\end{equation}
\begin{equation}
\label{eq:cont2}
    \nabla \cdot \bsu' = 0,
\end{equation}
that is, conservation of momentum and continuity of the fluctuating components. Notice that here the right-hand side of \eqref{eq:momentum3} has been condensed into a forcing term $\bsf'$ that represents the effect of the fluctuations about the mean state of the nonlinear terms, and can be regarded in this context as an exogenous input to a linear system comprising of the remaining terms. Wall-bounded parallel flows with a mean/base flow in the streamwise and spanwise dimensions $\bsU(y)=[U(y),0,W(y)]^T$, 
admit a transformation of variables from a primitive  reference $\{u',v',w',p'\}$ towards a reference in terms of the wall-normal velocity $v'$ and vorticity $\eta'$ (where $\eta' = \partial u'/\partial z-\partial w'/\partial x$), without loss of generality (i.e.~resulting in the Orr-Sommerfeld and Squire equations). This formulation is proven to be equally informative \citep{moarref2013channels,rosenberg2018efficient,mcmullen2020interaction} for resolvent analysis of planar flows. In this reference, the no-slip and no-penetration conditions translate into $v'(y=0)=v'(y=2h)=0$, $\pp_y v'(y=0)=\pp_y v'(y=2h)=0$ and $\eta'(y=0)=\eta'(y=2h)=0$, where $h$ represents the semi-height of the domain in the wall-normal dimension. Throughout this paper, the location of the no-slip walls will coincide with $y=0$ and $y=2h$. This transformation is achieved according to the process described in \citet{Schmid_Henningson}, while including a spanwise component of the mean/base flow, $W$. Note that while this spanwise mean component is included in the derivation, it will only be nonzero for the configuration presented in \S\ref{sec:results3Dchannel}. The resulting wall-normal formulation of the conservation laws shown in \eqref{eq:momentum3}--\eqref{eq:cont2} is formed by the following two equations
\begin{equation}
\label{eq:wallNormalVel}   
\left[\left(\pp_t+U\pp_x+W\pp_z\right) \nabla^2-\pp_y^2 U\pp_x-\pp_y^2 W\pp_z-\frac{1}{\Rey}\nabla^4\right] v'=f'_v,
\end{equation}
\begin{equation}
\label{eq:wallNormalVort}
    \left(\pp_t+U\pp_x+W\pp_z-\frac{1}{\Rey}\nabla^2\right)\eta'+\left(\pp_y U\pp_z -\pp_y W\pp_x\right)v'=f'_\eta.
\end{equation}
Assuming that the system is homogeneous in the temporal dimension and the streamwise and spanwise directions, we introduce assumed solutions of the form
\begin{subequations}
\label{eq:fourier}
\begin{eqnarray}
    v'(\bsx,t)=\hat{v} (y)\exp{[{i}(k_x x+k_z z-\omega t)]},    \\
    \eta'(\bsx,t)=\hat{\eta} (y)\exp{[{i}(k_x x+k_z z-\omega t)]},\\
    \bsf'(\bsx,t)=\hat{\bsf} (y)\exp{[{i}(k_x x+k_z z-\omega t)]},
\end{eqnarray}
\end{subequations}
where $k_x$ and $k_z$ denote the streamwise and spanwise wavenumbers, respectively, and $\omega$ denotes temporal frequency. Substituting these assumed solutions in \eqref{eq:wallNormalVel}--\eqref{eq:wallNormalVort} gives the following system of equations,
\begin{equation}
\label{eq:wallNMatrix3}
(-i\omega
\mathsfbi{M}+
\mathsfbi{L})
\begin{pmatrix}
\hat{v} \\
\hat{\eta}
\end{pmatrix}
=
\begin{pmatrix}
\hat{f}_{v} \\
\hat{f}_{\eta}
\end{pmatrix},
\end{equation}
where 
\begin{equation}
\mathsfbi{M}=
\begin{pmatrix}
\hat{\Delta} & 0\\
0& \mathsfbi{I}
\end{pmatrix} 
\text{ and }
\mathsfbi{L}=
\begin{pmatrix}
\mathcal{L}_{os} & 0 \\
i k_z \pp_yU-ik_x \pp_yW & \mathcal{L}_{sq}
\end{pmatrix}.
\end{equation}
The modified Laplacian operator $\hat{\Delta}=\pp_{yy}-(k_x^2+k_z^2)$ is introduced for simplicity, and
\begin{equation}
\label{eq:Los}    
\mathcal{L}_{os}=\left(ik_xU+ik_zW\right) \hat{\Delta}-\frac{1}{\Rey}\hat{\Delta}^2-ik_x\pp_y^2U-ik_z\pp_y^2W,
\end{equation}
\begin{equation}
\label{eq:Lsq}
    \mathcal{L}_{sq}=i k_x U+ik_zW-\frac{1}{\Rey} \hat{\Delta},
\end{equation}
represent the Orr-Sommerfeld (OS) and Squire (SQ) operators, respectively. 
Premultiplying both sides of equation \eqref{eq:wallNMatrix3} by $\mathsfbi{M}^{-1}$ and solving for the state $[\hat{v}(y),\hat{\eta}(y)]^T$ gives
\begin{equation}
\label{eq:wallNResolvent}
\begin{pmatrix}
\hat{v} \\
\hat{\eta}
\end{pmatrix}
=(-i\omega \mathsfbi{I}
+
\mathcal{L})^{-1}
\begin{pmatrix}
\hat{g}_{v} \\
\hat{g}_{\eta}
\end{pmatrix}:=\mathcal{H}_{\bomega} 
\begin{pmatrix}
\hat{g}_{v} \\
\hat{g}_{\eta}
\end{pmatrix},
\end{equation}
where
\begin{equation}
\mathcal{L}=
\mathsfbi{M}^{-1}\mathsfbi{L}
\text{ and }
\begin{pmatrix}
\hat{g}_{v} \\
\hat{g}_{\eta}
\end{pmatrix}=\mathsfbi{M}^{-1}
\begin{pmatrix}
\hat{f}_{v} \\
\hat{f}_{\eta}
\end{pmatrix}.
\end{equation}
This transfer function $\mathcal{H}_{\bomega}$ is denoted as the resolvent operator, in analogy to the definition introduced for a general linear system in \eqref{eq:resolventDynSys} when $z$ becomes $z=-i\omega$. Note that the resolvent operator is dependent on the triad $\{\omega,k_x,k_z\}$, but for the sake of readability this dependence is indicated by the subscript $\bomega$. Expanding the terms in accordance to the derivation described in \citet{rosenberg2018efficient} gives
\begin{equation}
\label{eq:wallNResolventInverse}
\begin{pmatrix}
\hat{v} \\
\hat{\eta}
\end{pmatrix}= \mathcal{H}_{\bomega} \begin{pmatrix}
\hat{g}_v \\
\hat{g}_\eta
\end{pmatrix}:=
\begin{pmatrix}
\mathcal{H}_{vv} & 0 \\
-\mathcal{H}_{\eta\eta} (i k_z \pp_yU-ik_x\pp_yW)\mathcal{H}_{vv} & \mathcal{H}_{\eta\eta}
\end{pmatrix}
\begin{pmatrix}
\hat{g}_v \\
\hat{g}_\eta
\end{pmatrix},
\end{equation}
where the scalar operators are defined as
\begin{equation}
\label{eq:hvv}
    \mathcal{H}_{vv}=(-i \omega+\hat{\Delta}^{-1}\mathcal{L}_{os})^{-1},
\end{equation}
\begin{equation}
\label{eq:hetaeta}
    \mathcal{H}_{\eta\eta}=(-i \omega+\mathcal{L}_{sq})^{-1}.
\end{equation} 

Notice that the formulation in wall-normal variables enables the study of the dynamical properties of each of the variables independently. Nevertheless, it is possible to define a direct transformation of the resolvent operator, as well as the resolvent modes, from wall-normal velocity and vorticity $\{v,\eta\}$ to primitive velocity variables $\{u,v,w\}$ (and vice versa), according to the mapping that was introduced in \citet{Meseguer2003Trefethen} and further developed and applied in \citet{jovanovic2005componentwise,mckeon2010resolvent,moarref2013channels,sharma2017scaling}. The cited transformation recasts the response and forcing modes in primitive variables as follows
\begin{equation}
    \boldsymbol{\Psi}_{(u,v,w)} = \mathsfbi{C} \boldsymbol{\Psi}_{(v,\eta)}, 
\end{equation}
and
\begin{equation}
    \boldsymbol{\Phi}_{(u,v,w)} = (\mathsfbi{M}^{-1}\mathsfbi{B})^{-1} \boldsymbol{\Phi}_{(v,\eta)}.
\end{equation}
Here
\begin{equation}
\label{eq:Ctransf}
\mathsfbi{C} = \frac{1}{k_\perp^2} 
\begin{pmatrix}
ik_x\partial_y & -ik_z \\
k_\perp^2  & 0 \\
ik_z\partial_y & ik_x 
\end{pmatrix},
\end{equation}
and 
\begin{equation}
\label{eq:Btransf}
\mathsfbi{B} =  
\begin{pmatrix}
-ik_x\partial_y & -(k_x^2+k_z^2) &-ik_z\partial_y \\
ik_z & 0 & -ik_x  
\end{pmatrix},
\end{equation}
represent the input and output matrices, respectively. 

\subsection{Space-time resolvent analysis}
\label{sec:spaceTimeResolvent}
Here, we present a form of resolvent analysis that is applicable to time-varying systems. This generalisation is achieved by limiting the assumed directions of homogeneity to the streamwise and spanwise spatial dimensions. Thus, in this formulation, both components of the mean state $\bsU=[U,0,W]^T$ are also assumed to be temporally-dependent, and we write a generalised instantaneous state $\bsu$ as
\begin{equation}
    \bsu(\bsx,t) = \bsU(\bsx,t)+\bsu'(\bsx,t).
\end{equation}
In analogy to the trajectories presented in \eqref{eq:fourier} we let the solutions take the following form
\begin{subequations}
\label{eq:fourierT}
\begin{eqnarray}
    v'(\bsx,t)=\hat{v} (y,t)\exp{[{i}(k_x x+k_z z)]},    \\
    \eta'(\bsx,t)=\hat{\eta} (y,t)\exp{[{i}(k_x x+k_z z)]},\\
    \bsf'(\bsx,t)=\hat{\bsf} (y,t)\exp{[{i}(k_x x+k_z z)]}.
\end{eqnarray}
\end{subequations}
Notice the more general dependence of these trajectories on both $y$ and $t$, allowing the solutions to adopt any sort of temporal function. Substituting these spatio-temporal solutions in the governing equations in wall-normal formulation in \eqref{eq:wallNormalVel}--\eqref{eq:wallNormalVort}, and solving for the current state $[\tilde{v}(y,t),\tilde{\eta}(y,t)]^T$ provides the following definition of the space-time resolvent operator $\mathcal{H}_{\boldsymbol{t}}$ 
\begin{equation}
\label{eq:wallNResolventTime}
\begin{pmatrix}
\hat{v} \\
\hat{\eta}
\end{pmatrix}= \mathcal{H}_{\boldsymbol{t}} \begin{pmatrix}
\hat{g}_v \\
\hat{g}_\eta
\end{pmatrix}:=
\begin{pmatrix}
\tilde{\mathcal{H}}_{vv} & 0 \\
-\tilde{\mathcal{H}}_{\eta\eta} (i k_z \pp_yU-ik_x\pp_yW)\tilde{\mathcal{H}}_{vv} & \tilde{\mathcal{H}}_{\eta\eta}
\end{pmatrix}
\begin{pmatrix}
\hat{g}_v \\
\hat{g}_\eta
\end{pmatrix},
\end{equation}
and the modified scalar operators are given by 
\begin{equation}
\label{eq:hvvMod}
    \tilde{\mathcal{H}}_{vv}=(\mathsfbi{D}_t+\hat{\Delta}^{-1}\mathcal{L}_{os})^{-1},
\end{equation}
\begin{equation}
\label{eq:hetaetaMod}    \tilde{\mathcal{H}}_{\eta\eta}=(\mathsfbi{D}_t+\mathcal{L}_{sq})^{-1}.
\end{equation} 
Here, $\mathsfbi{D}_t$ represents a generalised discrete time-differentiation operator, and the subscript in $\mathcal{H}_{\boldsymbol{t}}$ represents the triad $\boldsymbol{t}=\lbrace t,k_x,k_z\rbrace$. Notice that definitions \eqref{eq:hvvMod} and \eqref{eq:hetaetaMod} have the symbol $(\tilde{\cdot})$ to emphasise on the temporal dependence and disambiguate from \eqref{eq:hvv} and \eqref{eq:hetaeta}.
 As before, resolvent analysis proceeds by taking an SVD of the associated resolvent operator, $\mathcal{H}_t$. The leading resolvent forcing and response modes satisfy the same optimisation problems described in equations \ref{eq:sigopt},\ref{eq:phiopt}-\ref{eq:psiopt}, though now the norm is computed over both space and time, so that 
 \begin{equation}
 \label{eq:spacetime2norm}
    \|\bm\psi(\bsx,t) \|_2 = \left(\int_{t\in \Omega_t}\int_{\bsx\in \Omega_{\bsx}} \left|\bm\psi(\bsx,t)\right|^2 \text{d}\bsx\text{d}t\right)^{1/2},
\end{equation}
where $\Omega_t$ denotes the temporal domain under consideration.

While the theory of this generalization of resolvent analysis is straightforward, 
it does come with a potential increase in  the computational cost. Upon discretization, the size of the matrix representation of the resolvent operator is increased by a factor of the number of timesteps, $N_t$, in both the row and column dimensions with respect to the space-only resolvent operator defined in \eqref{eq:wallNResolvent}. For a case where one spatial dimension ($y$) is discretised, this means that the total size is $(2N_y N_t \times 2N_y N_t)$ and each of the block-elements is $(N_y N_t \times N_y N_t)$, where $N_y$ and $N_t$ are the number of discretisation points in the space and time dimensions, respectively. Note that the space-only formulation is constituted by a resolvent operator of total size $(2N_y \times 2N_y)$ and with block-elements of size $(N_y \times N_y)$. This increase is due to the fact that each of the entries of the operator $\mathsfbi{D}_t$ corresponds to a temporal instance of a given spatial location in the wall-normal axis. For the purposes of this study, however, this computational cost remains feasible.

Note that in order to disambiguate between the space-only modes and the space-time modes, the symbol $(\hat{\cdot})$ will be used to denote the space-only modes in \S\ref{sec:results}. 

\subsection{Sparse resolvent analysis}
\label{sec:sparseResolvent}

The theory presented in \S\ref{sec:pseudo} formulates the finding the leading resolvent modes and corresponding gain as an optimization problem (e.g.~\eqref{eq:sigopt}) in terms of the spatial $l_2$ norm of forcing and response modes (defined in \eqref{eq:space2norm}). Similarly, the leading resolvent modes for the space-time resolvent formulation described in \S\ref{sec:spaceTimeResolvent} is formulated using the norms computed over both the spatial and temporal domains \eqref{eq:spacetime2norm}. 

Such optimization problems involving the $l_2$ norm are ubiquitous across a broad range of methods, and arise naturally for methods based on the SVD. However, it is possible to modify such optimization problems such that their solution has different characteristic features.

Here, we introduce a variant of resolvent analysis that seeks to achieve localisation or sparsity, while also desiring the large energy amplification levels that are obtained in the standard resolvent formulation. This is achieved by incorporation of the variant of sparse principal component analysis (PCA) described in \citet{hein2010inverse}. Similar approaches are discussed in ~\citet{jolliffe2003modified,zou2006sparse,sigg2008expectation,journee2010generalized,zou2018selective}. 

Sparsity is promoted through the incorporation of an additional term in the form of the $l_1$-norm, which produces the following minimization problem
\begin{equation}
\label{eq:minsvdsparse}
    \boldsymbol{\psi}_1 =  \argmin_{\boldsymbol{\psi}} \frac{(1-\alpha)\|\boldsymbol{\psi}\|_2+ \alpha\|\boldsymbol{\psi}\|_1 }{\|\mathcal{H}^*\boldsymbol{\psi}\|_2}.
\end{equation}
Here, the sparsity parameter $\alpha\in [0,1]$ determines the number of nonzero elements in $\boldsymbol{\psi}_1$. Moreover, the sparsest solution will be retrieved when $\alpha=1$; while the case where $\alpha=0$ will give the least sparse outcome and will in fact match the result given by \eqref{eq:sigopt}. 

Notice that the numerator in \eqref{eq:minsvdsparse} is a convex function. The solution to this optimisation problem is achieved by reformulating it as a nonlinear eigenproblem that enables the use of an inverse power method to find its optimum \citep{hein2010inverse}. In practice, this method often produces solutions with sharp gradients, where some of the entries change drastically from zero to nonzero values. In order to produce coherent structures that resemble observable mechanisms, these solutions are regularised using the resolvent operator while maintaining sparsity. Here, we refer to the response modes computed on the first step as ``raw" modes, and use a superscript to disambiguate from the regularised or updated modes. The full collection of steps that produces the sparse leading forcing $\boldsymbol{\phi}_1$ and response $\boldsymbol{\psi}_1$ resolvent modes are described in algorithm \ref{alg:sparseRA}. 
\begin{algorithm}
    \caption{Simplified sparse resolvent analysis}\label{alg:sparseRA}
    \begin{algorithmic}[1]
        \vspace{0.05cm}
        \Statex \textbf{Input}: Resolvent operator $\mathcal{H}$ and sparsity parameter $\alpha$
        \vspace{0.05cm}
        \State Compute raw sparse response modes $\boldsymbol{\psi}^{raw}_1$ by solving \eqref{eq:minsvdsparse}
        \vspace{0.05cm}
        \State Compute corresponding forcing modes via
        \begin{equation}
            \label{eq:forcing}
             \boldsymbol{\phi}_1 = \frac{\mathcal{H}^*\boldsymbol{\psi}^{raw}_1}{\| \mathcal{H}^*\boldsymbol{\psi}_1^{raw}\|_2}
        \end{equation}
        \vspace{0.05cm}
        \State Compute updated (regularised) response modes $\boldsymbol{\psi}_1$ via
        \begin{equation}
            \label{eq:responseupdated}
            \boldsymbol{\psi}_1 = \frac{\mathcal{H}\boldsymbol{\phi}_1}{\| \mathcal{H}\boldsymbol{\phi}_1\|_2}
        \end{equation}
        \vspace{0.05cm}
        \State Compute corresponding singular values via
        \begin{equation}
        \label{eq:sparsesig}
            \sigma_1 = \| \mathcal{H}\boldsymbol{\phi}_1\|_2
        \end{equation}
        \Statex \textbf{Output}: Regularised sparse response mode $\boldsymbol{\psi}_1$, sparse forcing mode $\boldsymbol{\phi}_1$, singular value $\sigma_1$.
\end{algorithmic}
\end{algorithm}

The forcing modes could potentially be updated by substituting the updated response modes in \eqref{eq:forcing}, although in practice there is not a significant difference between the updated and the first forcing modes. In addition, it is possible to compute higher-order resolvent modes within this framework using the deflation scheme described in \citet{buhler2014flexible}. According to this, the components that have already been identified are removed from the optimisation space before following the steps presented above. Moreover, observe that the method described here promotes sparsity on the response modes $\boldsymbol{\psi}$, although exchanging $\boldsymbol{\psi}$ with $\boldsymbol{\phi}$ and $\mathcal{H}$ with $\mathcal{H}^*$ would yield sparsity-promotion in the forcing modes. Lastly, the subscript in the resolvent operator was removed in this section in order to indicate that this methodology can be applied to both the spatial and space-time resolvent operators. For an explicit and expanded form of this algorithm describing how the solution to \eqref{eq:minsvdsparse} is obtained, refer to appendix~\ref{appA}. 

Notice that to preserve consistency in the cases presented here, we prescribe the value of the sparsity ratio $\gamma$, instead of the sparsity parameter $\alpha$, as an input to the algorithm. This value is defined as the number of nonzero entries divided by the number of total entries in $\boldsymbol{\psi}$. The relationship between the sparsity parameter $\alpha$ and the sparsity ratio $\gamma$ is further described in appendix~\ref{appA}. The advantage of prescribing an input value of the sparsity ratio $\gamma$ instead of the sparsity parameter $\alpha$ directly, is the fact that the algorithm gives control to the user in terms of the desired nonzero elements contained in the computed sparse modes. To some extent, it also allows the user to have control over the number of key features of the flow that are captured by the sparse mode. Note prescribing a set value of $\alpha$ instead requires careful adjustment to achieve the desired sparsity, since it could yield modes with a varying number of nonzero elements.

\section{Results}
\label{sec:results}
In this section, we present the application of the proposed framework on four different systems. After describing numerical details associated with the resolvent calculations in \S\ref{sec:numerics}, in \S\ref{sec:resultsChannel1d} we showcase the implementation of sparse resolvent analysis on a statistically-stationary turbulent channel flow. Here, we consider conventional resolvent analysis with a Fourier transform in time, but enable sparsity promotion in the spanwise direction. We then consider the space-time resolvent operator for this problem in \S\ref{sec:resultsChannel2d}, and show that the proposed method can produce temporally-localised modes for a statistically stationary flow. We next apply 
sparse and non-sparse space-time resolvent analysis to two non-statistically-stationary systems: a periodic turbulent Stokes boundary layer in \S\ref{sec:resultsStokesBL}, and a turbulent channel flow with sudden lateral pressure gradient in \S\ref{sec:results3Dchannel}.

\subsection{Numerical methods}
\label{sec:numerics}

The mean velocity profiles used for resolvent analysis are all obtained from direct numerical simulations (DNS).   These simulations use a staggered second-order finite difference scheme \citep{orlandi2000fluid}, with a fractional step method \citep{kim1985application} and third-order Runge-Kutta time-advancing scheme \citep{wray1990minimal}. Further details regarding the use and validations of these methods, and their application to the specific cases considered here, can be found in \citet{bae2018turbulence,bae2019dynamic,lozano2019characteristic}.


For resolvent analysis, the wall-normal direction is discretised using a Chebyshev collocation method. In the cases where the spanwise dimension is explicitly discretised, we use a Fourier discretisation scheme with periodic boundary conditions along the spanwise domain. Moreover, if homogeneity is not assumed in the temporal dimension, we adopt a Fourier discretisation scheme when the system is assumed to be statistically-stationary or time-periodic (i.e. \S\ref{sec:resultsChannel1d}--\S\ref{sec:resultsStokesBL}), and an explicit Euler finite-differentiation  scheme in the temporal dimension with Neumann boundary conditions at the boundaries in \S\ref{sec:results3Dchannel}. The corresponding differentiation operators for both Chebyshev and Fourier discretisations are defined according to the specifications given in \citet{weideman2000matlab}. The number of collocation points used for each of the examples considered in this work will be indicated in the corresponding section. In each case, we verify that the numerical resolution gives converged results, with details of these convergence studies given in appendix~\ref{appB}.

In the space-time implementations of the analysis showcased in this paper (i.e. \S\ref{sec:resultsChannel2d}--\S\ref{sec:results3Dchannel}), no-slip and no-penetration conditions are enforced at $y=0$ (lower wall), while free-slip and no-penetration conditions are enforced at $y=h$ (the channel centerline). The space-time analysis identifies modes that are sufficiently localised away from the mid-plane using the chosen set of parameters, allowing us to reduce the size of the numerical domain to the area below $y/h=1$.

\subsection{Spatially-sparse resolvent analysis of turbulent channel flow}
\label{sec:resultsChannel1d}
Here, we apply the sparse resolvent analysis methodology on a fully-developed turbulent channel flow, where we consider spatial (rather than temporal) sparsity. The mean velocity profile is obtained from direct numerical simulations at a friction Reynolds number of $\Rey_\tau=186$, defined as $\Rey_\tau=u_\tau h/\nu$ and the friction velocity, expressed as $u_\tau=\sqrt{\tau_w/\rho}$. The variables $\nu$ and $\rho$ represent the kinematic viscosity and density of the flow, respectively, the shear stress at the wall-boundary is denoted as $\tau_w$, and $h$ is the channel half-height. Hereafter, the superscript $(\cdot)^+$ denotes viscous (inner) units. In this reference, the velocities are scaled by the friction velocity $u_\tau$, and the lengthscale is given by $\nu/u_\tau$. In order to showcase the application of the sparse formulation of resolvent analysis introduced in \S\ref{sec:sparseResolvent}, we consider two configurations, both of which assuming homogeneity in the streamwise direction and the temporal dimension: a first case where the spanwise dimension is assumed to have an infinite extent, followed by a second implementation in which the spanwise dimension is limited to a finite periodic domain. 

The first configuration represents a one-dimensional analysis, where the chosen wavelengths in the streamwise $\lambda_x^+=1000$ and spanwise $\lambda_z^+=100$ directions correspond to the average size of the streaks and vortices that arise in the near-wall cycle \citep{jimenez1999autonomous}. The temporal frequency is fixed at $\omega=17.14$, where the temporal scale is normalised by the friction velocity  ($u_\tau$) and the half-height of the channel ($h$). The frequency is chosen to be that which yields the maximum resolvent gain at these wavelengths, which will best enable direct comparison with the following space-time resolvent results. Note that this frequency gives a critical layer at a location $y^+\approx 40$ inner units from the wall, which is further from the wall than typical near-wall streaks (which are characteristically found at $y^+\approx 15$). To avoid any misinterpretation, henceforth we will generally refer to the resulting structures coming from resolvent response modes (which tend to have amplitude peaks near the critical layer) simply as streamwise (rather than near-wall) streaks and vortices. Note also that the location of these structures could be moved closer to the wall either by reducing the choice of $\omega$ (which we find yields qualitatively similar results), or by introducing an eddy viscosity term to the resolvent formulation \citep{symon2023use}. In this application, the number of collocation points in the wall-normal direction is $N_y=101$. No-slip and no-penetration conditions are enforced for the velocity at the upper and lower boundaries.

\begin{figure}
\centering {
\vspace{-0.cm}
{\hspace*{-0.3cm}\includegraphics[width= 1\textwidth]{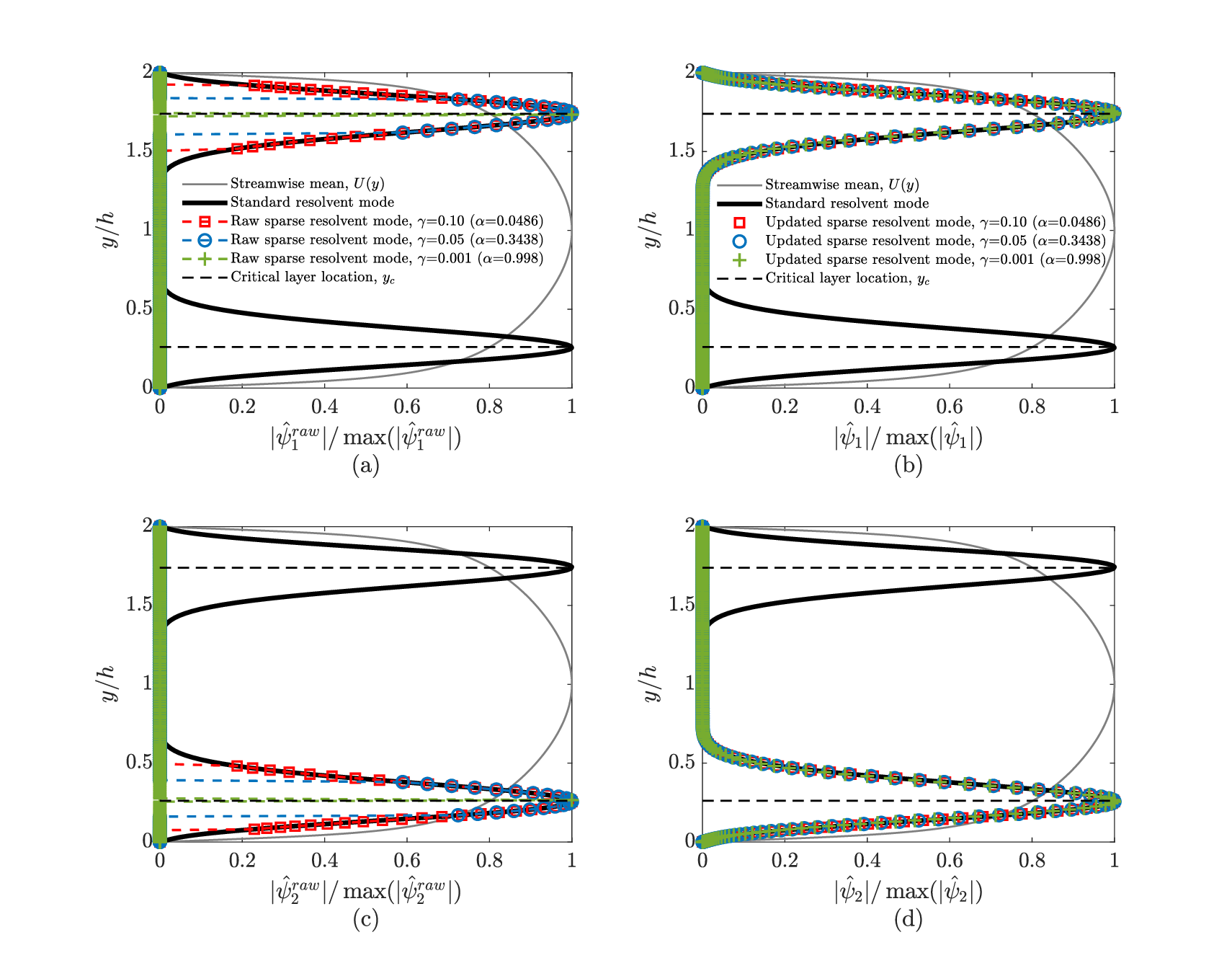}} }
\vspace{-0.15cm}
\caption{Streamwise components ($u$) of the first (\textit{a,b}) and second (\textit{c,d}) standard and sparse resolvent response modes computed for a turbulent channel flow with $\Rey_\tau=186$, $\lambda_x^+=1000$, $\lambda_z^+=100$ and temporal frequency $\omega=17.14$. Note that the standard modes (black) have been computed using \eqref{eq:psiopt}, while \eqref{eq:minsvdsparse} produced the raw sparse modes $\hat{\boldsymbol{\psi}}^{raw}$ (\textit{a,c}) from which we extract the adjusted modes $\hat{\boldsymbol{\psi}}$ (\textit{b,d}). The streamwise mean flow $U$ is added to all plots for reference. The black dashed lines denote critical layer locations.
}
\vspace{-0.1cm}
\label{fig:sparse1D}
\end{figure}

Results obtained from applying both standard and sparsity-promoting (in the wall-normal direction) resolvent  analysis are shown in figure~\ref{fig:sparse1D}, where we show the amplitude of the streamwise velocity component of the leading two resolvent response modes. For standard resolvent analysis, the chosen wavenumbers and frequency produces modes with two peaks, each centered near one of the two critical layer locations (where $U(y) = \omega/k_x$). The leading two singular values are essentially identical, and represent the fact that the two peaks are separated from each other, and can each represent structures that have an arbitrary phase shift between them. Mathematically, these modes are basis vectors for a subspace of dimension two, and thus an equally valid choice of basis for this subspace would be given by localised modes with one peak each. 

To compare to the results of sparse resolvent analysis, several solutions are computed for different sparsity ratios $\gamma$ via the optimisation problem posed in \eqref{eq:minsvdsparse}. The obtained raw sparse resolvent modes are shown in figures~\ref{fig:sparse1D}(\textit{a}) and ~\ref{fig:sparse1D}(\textit{c}) seem to be highly dependent on the corresponding value of $\gamma$, although they are again all located near one of the two critical layer locations. This dependence on the sparsity parameter subsides after regularizing the modes following algorithm \ref{alg:sparseRA}, as shown in figures~\ref{fig:sparse1D}(\textit{b}) and ~\ref{fig:sparse1D}(\textit{d}). We observe that these regularized modes each recover one of the two peaks identified by regular resolvent analysis, indicating that the sparse variant is finding basis elements for the leading resolvent subspace that are spatially sparse. The sparse singular values (computed via \eqref{eq:sparsesig}) are also consistent with the standard resolvent case, with $\sigma_1$ being about 1\% lower for the sparse case with $\gamma = 0.1$ ($\alpha = 0.0486$) in comparison with the non-sparse equivalent.

As the regularisation step appears to both remove the dependence on the sparsity parameter, and recovers the results for regular resolvent analysis for this example, we will exclusively use this regularised version for the remaining cases.  

We now consider the same turbulent channel flow, but instead of assuming a Fourier decomposition in the spanwise direction, we instead explicitly discretise this dimension, applying periodic boundary conditions with a spanwise extent $L_z$ twice the channel height (that is $L_z/h=4$). We use a Fourier basis in this spanwise dimension, with $N_z=92$ collocation points (with $N_y=101$ as before). We keep the same frequency and streamwise wavelength that was used in the one-dimensional analysis. This configuration is motivated by the fact that structures and correlations 
that are observed in wall-bounded turbulent flows typically exist only over a finite spanwise extent \citep{kim1987turbulence, hutchins2007evidence,dennis2011experimental,sillero2014turbulentBL,jimenez2018coherent}. While such localized structures could be represented by an appropriate combination of spanwise Fourier modes, this would be less efficient than a single-mode representation.

\begin{figure}
\centering {
\vspace{-0.cm}
{\hspace*{-1.1cm}\includegraphics[width= 1.15\textwidth]{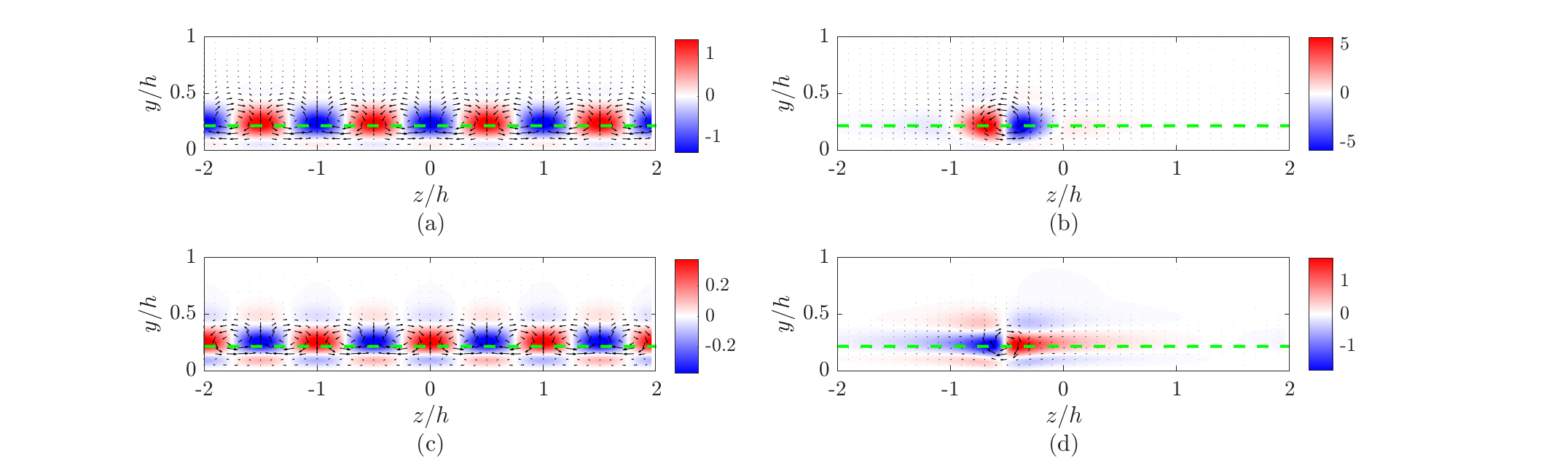}} }
\vspace{-0.5cm}
\caption{Real component of leading response modes $\hat{\boldsymbol{\psi}}_1$ (\textit{a-b}) and forcing modes $\hat{\boldsymbol{\phi}}_1$ (\textit{c-d}), for standard (\textit{a,c}) and sparse (\textit{b,d}) with $\gamma = 0.001$ ($\alpha = 0.953$) resolvent analysis, applied to turbulent channel flow at $\Rey_\tau = 186$, with a finite periodic domain in the spanwise direction defined by $z/h \in \lbrace-2,2\rbrace$ and a frequency $\omega=17.14$. Contours of streamwise velocity are shown, with arrows indicating the velocity in the spanwise ($z$) and wall-normal ($y$) directions. Green dashed lines indicate critical layer locations. The streamwise wavelength is $\lambda_x^+$ = 1000.
}
\label{fig:spanwiseChann}
\end{figure}

Figure~\ref{fig:spanwiseChann} shows the leading resolvent forcing and response modes obtained from standard and sparse resolvent analyses, visualised in the $y-z$ plane. Note that the contours represent the streamwise component ($u$) of the modes, which correspond to streamwise streaks of fast- and slow-moving regions. The vector fields represent the wall-normal ($v$) and spanwise ($w$) velocity components of the modes, which here form streamwise-aligned vortical structures. Since the system is homogeneous in the spanwise dimension, the standard resolvent modes (figure~\ref{fig:spanwiseChann}(\textit{a,c})) give Fourier modes in this direction. The response mode consists of alternating slow- and fast-moving streamwise streaks, with streamwise vortices located between each streak. 
The wall-normal profile of these modes is close to matching those from the one-dimensional analysis shown in figure \ref{fig:sparse1D}, where the identified spanwise wavelength for the leading mode is slightly different from that selected in the one-dimensional analysis shown in figure \ref{fig:sparse1D}. The configuration of these streamwise streaks is consistent with the the lift-up mechanism \citep{landahl1975wave,landahl1980note}, through which streamwise vortices lead to the formation of streamwise streaks by transporting slow-moving fluid away from the wall, and vice-versa. 

The sparse resolvent modes shown in figure~\ref{fig:spanwiseChann}(\textit{b,d}) contain a spatially-localised unit of the periodic  structures identified by standard resolvent in figure~\ref{fig:spanwiseChann}(\textit{a,c}). In particular, the sparse response mode consists of a primary central vortex surrounded by a fast and slow streak, flanked by lower-amplitude secondary vortices and streaks. Note that while we are only showing the lower half of the domain, on the upper half the same structure is present for standard resolvent, but not for the sparse equivalent (again consistent with figure \ref{fig:sparse1D}) with $\gamma = 0.001$ ($\alpha = 0.953$). While not shown here, higher-order sparse resolvent modes consist of repetitions of this localized structure both on the upper wall, and translated in the spanwise direction (with the spanwise location of the leading mode being arbitrary). Note also that the relative phase of the forcing and response modes is consistent between the regular and sparse modes. In terms of the energy content of these structures, the first non-sparse singular value is about 1.178 times larger than its sparse counterpart.

\begin{figure}
\centering {
\vspace{-0.cm}
{\hspace*{-.3cm}\includegraphics[width= 1\textwidth]{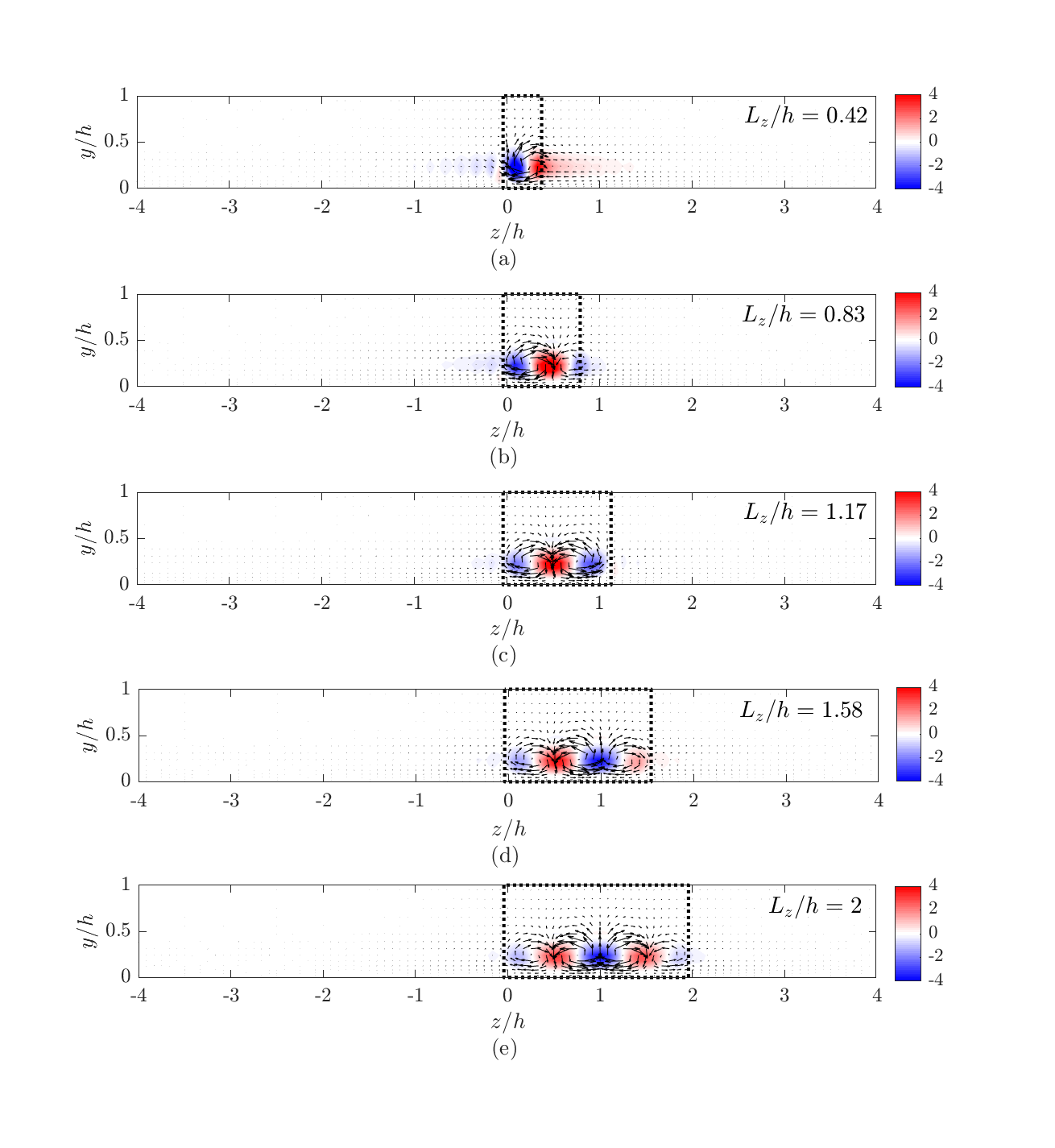}} }
\vspace{-1cm}
\caption{Real component of leading response modes $\hat{\boldsymbol{\psi}}_1$ for standard resolvent analysis, applied to turbulent channel flow at $\Rey_\tau = 186$, with a finite periodic domain in the spanwise direction restricted to $L_z/h \in \{0.42,0.83,1.17,1.58,2 \}$ with $\omega=17.14$ and $\lambda_x^+$ = 1000. Contours of streamwise velocity are shown, with arrows indicating the velocity in the spanwise ($z$) and wall-normal ($y$) directions.}
\label{fig:modes_restricted}
\vspace{-0cm}
\end{figure}

To further explore the relationship between the sparse and non-sparse modes for this configuration, we show infigure~\ref{fig:modes_restricted} modes computed using the standard version of resolvent analysis, but the region where amplification is measured is restricted in the spatial domain along the $z$-axis (while maintaining a domain of infinite extent in the streamwise direction). This is achieved by modifying the weight function associated with the resolvent operator to only consider amplification at spatial locations within the indicated regions. Note that in this case, the spanwise domain is defined over a larger region $-4 \leq z/h \leq +4$ to further emphasise this localization. As in the previous results for both sparse and non-sparse analyses, the identified modes consist of alternating slow- and fast-moving streamwise streaks, with streamwise  vortices between each pair of streaks. The size of these structures is approximately the same once the window reaches a width $L_z \gtrapprox h$, with the
number of such structures present dependent on the width of the window.
Restricting the spanwise domain also reduces energy amplification associated with these structures, in comparison with the Fourier mode extending across the entire domain width. 

\begin{figure}
\centering {
\vspace{-0.cm}
{\hspace*{-0.2cm}\includegraphics[width= 0.7\textwidth]{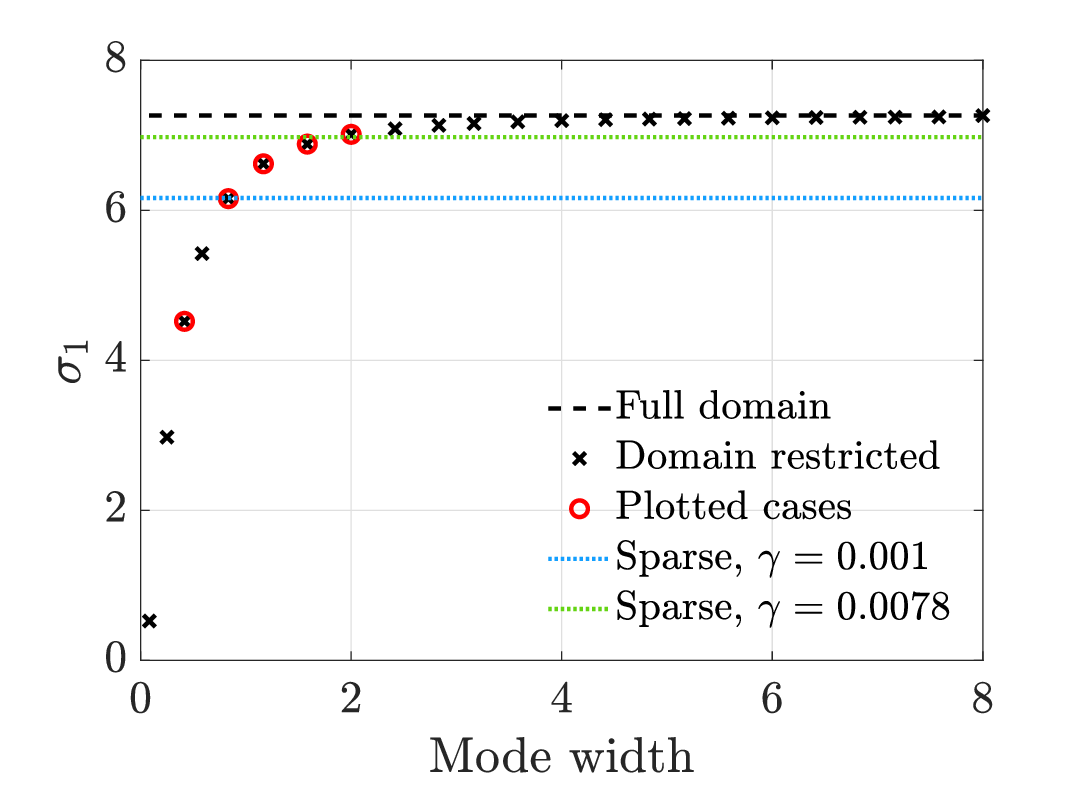}} }
\vspace{0.2cm}
\caption{Leading singular values of the resolvent operator for a restricted spanwise domain of length $0.08 \leq L_z/h \leq 8$. The values corresponding to the modes compiled in figure~\ref{fig:modes_restricted} are indicated with red markers. Note that the leading singular value of the non-sparse and sparse resolvent modes depicted in figure~\ref{fig:spanwiseChann} are also indicated with dashed black and blue dotted lines, respectively, for reference. In addition, the leading singular value of the sparse mode  retrieved for $\gamma=0.0078$ ($\alpha=0.625$) is indicated with a green dotted line.}
\label{fig:singVals_restricted}
\vspace{-0.0cm}
\end{figure}

Figure~\ref{fig:singVals_restricted} shows the leading singular value of the resolvent modes as a function of the mode width $(L_z/h)$,  along with the singular values associated with the the sparse and non-sparse modes depicted in figure~\ref{fig:spanwiseChann} (blue dotted line). As the mode width increases, we observe that the singular value increases rapidly for small $L_z$, and then becomes more gradual as the singular value approaches that obtained using the full domain. When $L_z = 0.83h$ (figure~\ref{fig:modes_restricted}(\textit{b})), we observe a mode structure similar to that found for the sparse mode, with a pair of central positive and negative streamwise streaks and a central streamwise vortex sitting between the streak pair. In figure~\ref{fig:singVals_restricted} it is observed that this case is located near the elbow of the amplification against mode width curve, with a singular value approximately matching that obtained for the sparse mode (blue dotted line). This suggests that the sparse mode is identifying a structure that finds an appropriate trade-off between amplification and sparsity, by identifying a structure that is much more localized that the original spanwise Fourier mode, while still having a comparable singular value. 
To show the effect that the choice of sparsity parameter has on this trade-off between amplification and localization, we also show in figure~\ref{fig:singVals_restricted} (green dotted line) the sparse singular value for a larger sparsity ratio of $\gamma=0.0078$ ($\alpha=0.625$). This amplification level intersects the $\sigma_1$ versus mode width curve at a larger mode width. The associated sparse resolvent response mode, while not plotted, was found to be similar to the configuration of streamwise streaks and vortices identified in figure~\ref{fig:modes_restricted}\textit{(d)}, which is near this intersection point.

\subsection{Space-time resolvent analysis of a turbulent channel flow}
\label{sec:resultsChannel2d}
In this section, we demonstrate the implementation of the non-sparse and sparse space-time variants of resolvent analysis on the statistically-stationary turbulent channel flow considered in \S\ref{sec:resultsChannel1d}. Here, we do not perform a Fourier transform in the temporal dimension, and instead implement the space-time formulation of resolvent analysis introduced in \eqref{eq:wallNResolventTime}. All the parameters adopted in the analysis conducted in \S\ref{sec:resultsChannel1d} are also adopted here (except for the frequency, which is no longer specified), and time is nondimensionalised by the friction velocity, $u_\tau$, and the half-height of the channel, $h$.

\begin{figure}
\centering {
\vspace{-0.cm}
{\hspace*{-1.cm}\includegraphics[width=1.08\textwidth]{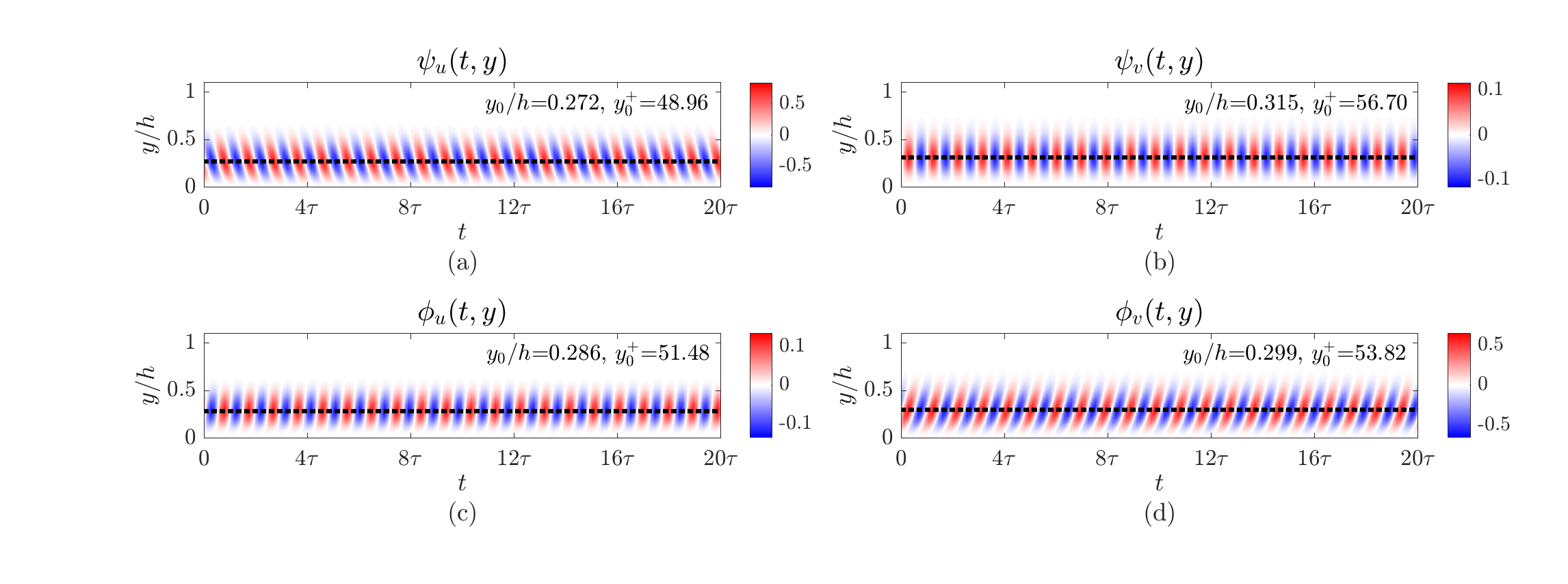}} }
\vspace{-0.8cm}
\caption{Real component of space-time leading response $\boldsymbol{\psi}_1$ (\textit{a-b}) and forcing $\boldsymbol{\phi}_1$ (\textit{c-d}) modes of the resolvent operator in a turbulent channel flow with $\Rey_\tau=186$, and $\lambda_x^+=1000$, $\lambda_z^+=100$ for a time horizon $\tau_{tot}=20\tau=20(2\pi/\omega)$. Horizontal dotted lines in (\textit{a-d}) indicate the $y$-location where each mode component achieves its maximum amplitude ($y_0/h$). Subplots (\textit{a,c}) show the streamwise velocity component ($u$), while (\textit{b,d}) give the wall-normal component ($v$).
}
\label{fig:spaceTime_turbchannel}
\vspace{-0.0cm}
\end{figure}

First, the streamwise ($u$) and wall-normal velocity ($v$) components of the resolvent modes obtained from the non-sparse space-time analysis are shown in figure~\ref{fig:spaceTime_turbchannel} with a time horizon $\tau_{tot}=20\tau$ that spans over 20 cycles of length $\tau=2\pi/\omega=0.3659$, where $\omega = 17.14$ is the same frequency used in \S\ref{sec:resultsChannel1d}-\ref{sec:resultsChannel2d}. The number of collocation points is $N_y=101$ in the wall-normal direction and $N_t=551$ in the temporal dimension. The location of maximum mode amplitude along the $y$-axis is indicated with a horizontal dotted line, and the value is indicated for each subplot with the symbol with $y_0/h$ and its equivalent in inner units $y_0^+$ for reference.

As expected, we observe that the leading spacetime resolvent mode are Fourier modes in time, with a frequency that matches that used in the previous analyses (i.e. the mode exhibits twenty periods over the time domain of length $20\tau$).

\begin{figure}
\centering {
\vspace{0cm}
{\hspace*{0cm}\includegraphics[width= 0.55\textwidth]{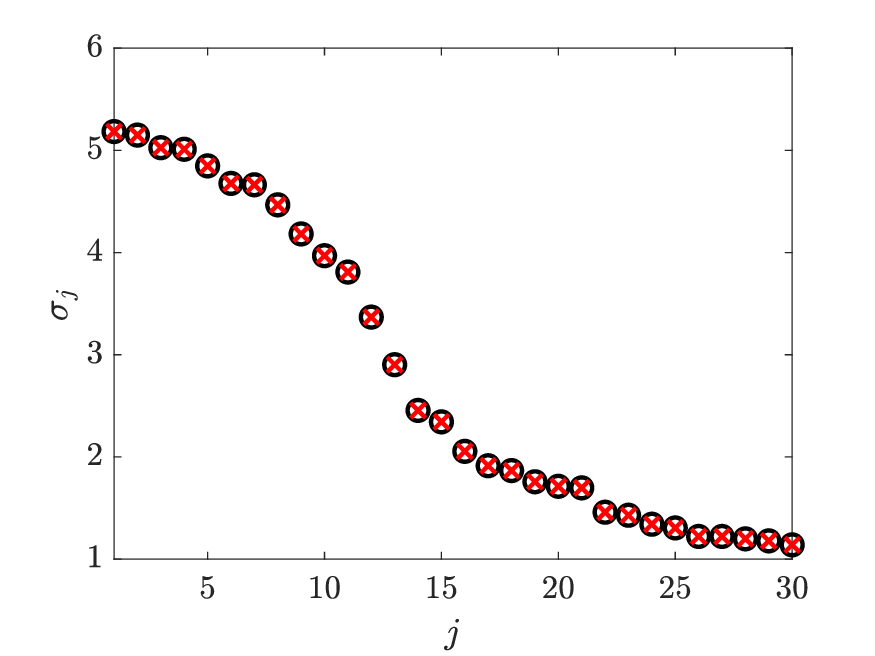}} }
\vspace{0.2cm}
\caption{Leading $30$ singular values of the space-time resolvent (black) and singular values of the space (standard) resolvent computed at $\omega_n=n(2\pi/\tau_{tot})$ with $-100\leq n \leq 100$ sorted in descending order (red) for a turbulent channel flow with $\Rey_\tau=186$, $\lambda_x^+=1000$, $\lambda_z^+=100$ with a time horizon (for the space-time variant) $\tau_{tot}=20\tau=20(2\pi/\omega)$, where $\omega=17.14$.
}  
\label{fig:spaceTime_channel_singvals}
\end{figure}

To further demonstrate that the spacetime variant  gives equivalent results to standard resolvent analysis, in figure~\ref{fig:spaceTime_channel_singvals} we compare the leading 30 singular values from both versions, where in the standard resolvent analysis we compile the results across all frequencies that are permissible when using this temporal domain (i.e.~ $\omega=2\pi n/\tau_{tot}$ with $n \in \mathbb{Z}$).

\begin{figure}
\centering {
\vspace{-0.cm}
{\hspace*{-1.cm}\includegraphics[width= 1.08\textwidth]{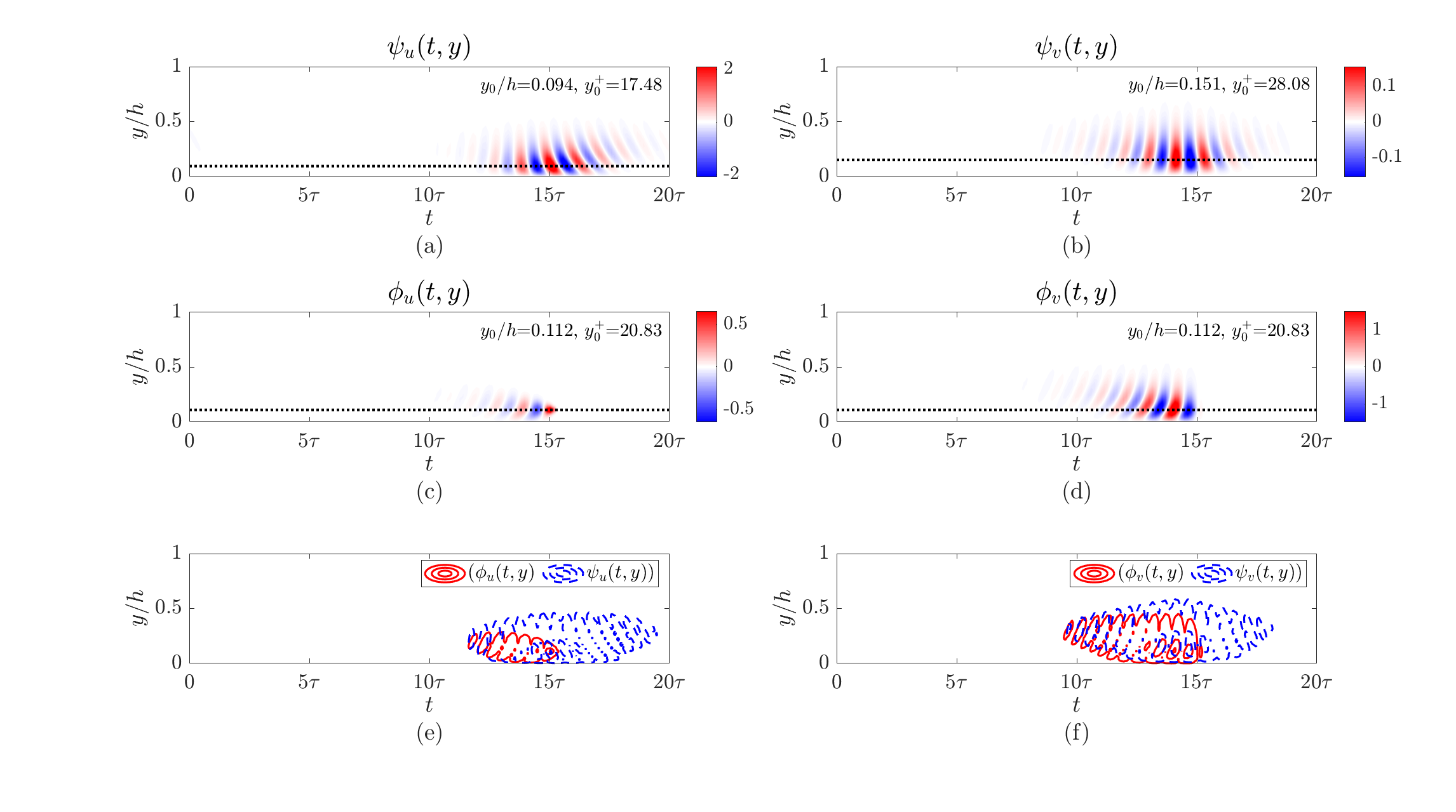}} }
\vspace{-0.8cm}
\caption{Real component of space-time sparse leading response $\boldsymbol{\psi}_1$ (\textit{a,b}) and forcing $\boldsymbol{\phi}_1$ (\textit{c,d}) modes of the resolvent operator in a turbulent channel flow with $\Rey_\tau=186$, and $\lambda_x^+=1000$, $\lambda_z^+=100$, computed with a sparsity parameter $\gamma=0.001$ ($\alpha = 0.094$), for a time horizon $\tau_{tot}=20\tau=20(2\pi/\omega)$.
Horizontal dotted lines in (\textit{a-d}) indicate the $y$-location where each mode component achieves its maximum amplitude ($y_0/h$). Subplots (\textit{e,f}) show contour levels of the absolute value of leading forcing (red) and response (blue) modes.}
\label{fig:spaceTime_turbchannel_sparse}
\vspace{-0.0cm}
\end{figure}

We now consider the spatio-temporally sparse variant of this analysis. The total time horizon is again set to $\tau_{tot}=20\tau$, which allows for the potential growth and decay of temporally-localised modes without the influence of the periodic boundary conditions. $N_y=101$ collocation points are used  in the wall-normal direction, and $N_t=501$ in the time dimension. 
Figure~\ref{fig:spaceTime_turbchannel_sparse} contains the streamwise and wall-normal components of the updated response and forcing modes with a sparsity ratio $\gamma=0.001$ ($\alpha = 0.883$). The location of maximum mode amplitude along the $y$-axis is again indicated with a horizontal dotted line. Notice how the analysis identifies structures that are sparse both in the spatial and temporal dimensions. 
Observe that the forcing modes in figure~\ref{fig:spaceTime_turbchannel_sparse}(\textit{c,d}) precede the response in figure~\ref{fig:spaceTime_turbchannel_sparse}(\textit{a,b}). The phase variation in time (corresponding to the temporal frequency) and wall-normal location of the modes approximately match those identified from the leading non-sparse space-time resolvent mode shown in figure~\ref{fig:spaceTime_turbchannel}. The sparse mode, however, identifies aspects of the physics that cannot be directly observed without sparsity-promotion (and thus localisation), such as the change in inclination angle and wall-normal location of the mode components with time. In particular, the inclination angle of structures tend to lean further backwards in the $y-t$ plane as time increases, which would correspond to an increased downstream tilt over time in the $x-y$ plane, consistent with the Orr mechanism \citep{orr1907stability}. 

We also note that the temporal locations of these time-localized resolvent mode components are consistent with the inherent causal structure of the time domain, in which the forcing precedes the response in the same way that an input to a dynamical system precedes an output. This temporal structure is also consistent with known physical mechanisms as well, with forcing in $v$ proceeding response in $v$ (figure~\ref{fig:spaceTime_turbchannel_sparse}\textit{f}) which in turn proceeds the (energetically-dominant) response in $u$ (figure~\ref{fig:spaceTime_turbchannel_sparse}\textit{e}), consistent with the lift-up mechanism. In terms of the energy content, for a fixed time horizon of $20\tau$, the leading non-sparse singular value is larger than the leading sparse one by a factor of 5.60. 

\begin{figure}
\centering {
\vspace{-0.cm}
{\hspace*{0cm}\includegraphics[width= 0.95\textwidth]{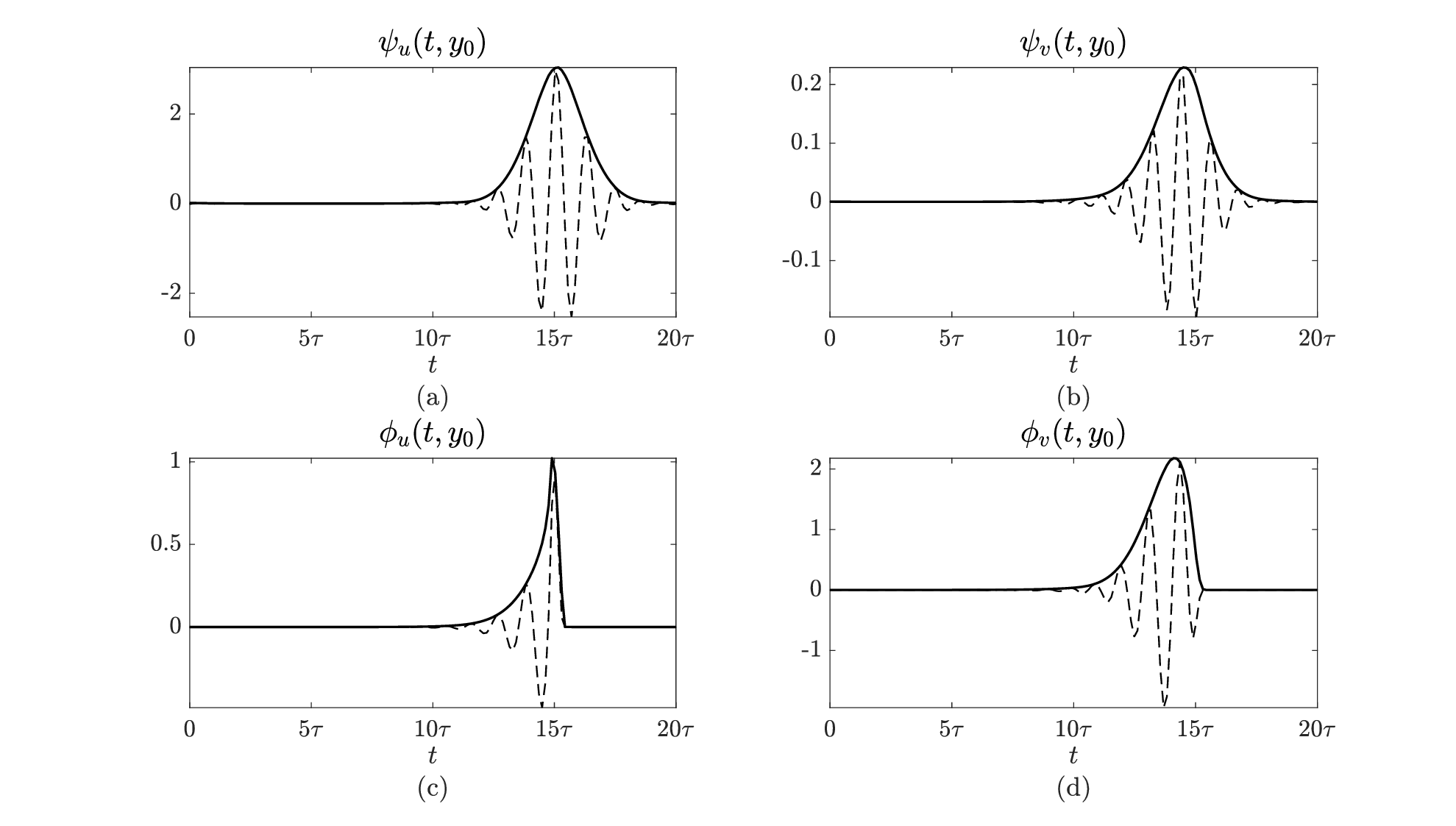}} }
\vspace{0cm}
\caption{Absolute (solid) and real (dashed) components of the temporal evolution of the sparse modes shown in figure~\ref{fig:spaceTime_turbchannel_sparse}(\textit{a-d}), at the $y$-locations of maximum amplitude ($y_0$). 
}
\label{fig:spaceTime_turbchannel_sparse_cross}
\end{figure}

To explore the temporal profiles of the components of these sparse space-time modes, figure~\ref{fig:spaceTime_turbchannel_sparse_cross} shows the cross-sections along the $t$-axis of the sparse modes in figure~\ref{fig:spaceTime_turbchannel_sparse} at the locations in $y$ where the amplitude of the signal finds its maximum. The temporal evolution of the forcing mode components seems to be more localised in these cross-sections, as they abruptly drop to zero near where the amplitude of the response modes finds its maximum value. The locations of the temporal peaks of these mode components is consistent with the discussion above related to figure~\ref{fig:spaceTime_turbchannel_sparse}, with the peak amplitude $\phi_v$ preceding that of $\psi_v$, which itself comes prior to the maximum in $\psi_u$. Quantitatively, the time duration between the peaks is on the same order of magnitude as $\tau$, with a time lag of approximately $0.41\tau$ between the peaks in the $v$ components of the forcing and response modes, and a further time lag of $0.68\tau$ between the peak amplitude of the responses in $v$ and $u$. This suggests that the energy transfer between these mode components occurs over a single oscillatory cycle. Note also that the phase lag in the between the $\psi_v$ and $\psi_u$ peaks approximately matches the phase difference between the oscillating component of these signals, as both are entirely real near their peak amplitudes. The shape of the streamwise and wall-normal components of the modes (first row in figure~\ref{fig:spaceTime_turbchannel_sparse_cross}) displays a Gaussian envelope with a nearly constant phase gradient and a uniform phase variation across all mode components. This could potentially suggest a natural choice of wavelet basis functions for a low order temporally-localised analysis of this flow \citep{ballouz2023wavelet,ballouz2024wavelet,Madhusudanan2024wavelet}. Moreover, the particular shape of the response modes could enable the implementation of wavepacket pseudomode theory to approximate these response modes \citep{obrist2010algebraically,obrist2011algebraically,mao2011continuous,edstrand2018parallel,dawson2019shape,dawson2020ijhff}.

\subsection{Space-time resolvent analysis of a turbulent Stokes boundary layer}
\label{sec:resultsStokesBL}
\begin{figure}
\centering {
\vspace{0.2cm}
{\hspace*{-0cm}\includegraphics[width= 1\textwidth]{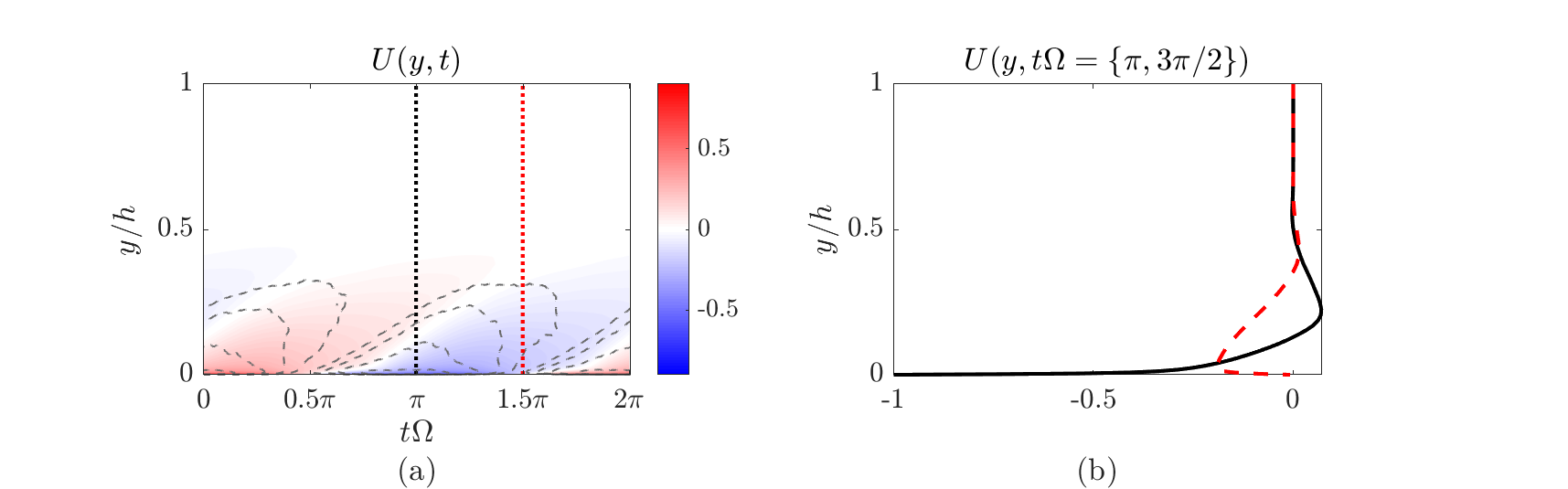}} }
\vspace{-0.4cm}
\caption{(a) Streamwise mean flow $U(y,t)$ (colourmap) and streamwise turbulence intensity $u_{rms}$ (grey dashed) of a turbulent Stokes boundary layer, with levels indicating \{20\%, 30\%, 50\%, 75\%, 95\%\} of its maximum value, over a full oscillating cycle; (b) time-instances of the turbulent mean at $t\Omega=\pi$ (black) and $t\Omega=3\pi/2$ (red).
}
\vspace{-0.0cm}
\label{fig:turbStokes_mean}
\end{figure}

In this section, we apply non-sparse and sparse space-time resolvent analysis to a turbulent Stokes boundary layer, with a time-periodic  mean velocity field as shown in figure~\ref{fig:turbStokes_mean}. This flow sits between two plates that oscillate synchronously with a velocity given by
\begin{equation}
\label{eq:turbMean}
    U_w(t) = U_0\cos{(\Omega t)},
\end{equation}
with $U_w$ denoting the velocity of the oscillating walls and $U_0$ representing a constant. 
There is no external pressure gradient, so the flow is entirely driven by the motion of the walls. The Reynolds number for this flow is defined as a function of the frequency of the oscillations $\Omega$, such that $\Rey_\Omega=U_0\delta_\Omega/\nu$, and $\delta_\Omega=\sqrt{2\nu/\Omega}$ is a length parameter denoting the boundary layer thickness. For the ensuing analysis, we choose $\Rey_\Omega=1500$ 
to ensure that the flow lies in the intermittently turbulent regime \citep{hino1976pipe,akhavan1991investigation,verzicco1996direct,vittori1998direct,costamagna2003coherent}. Note that while there appears to be little prior work performing linear analyses of this configuration in the turbulent regime, linear analysis of laminar pulsatile channel flow has been considered using Floquet analysis \citep{pier2017linear} and optimally time-dependent modes \citep{kern2021transient}.

The mean velocity profile and the root-mean-square (r.m.s.) of the fluctuating turbulent velocity components are obtained through direct numerical simulations as described in \S\ref{sec:numerics}. The size of the domain for the DNS is $6\pi\delta_\Omega \times 80\delta_\Omega \times 3\pi\delta_\Omega$ (in the streamwise, wall-normal, and spanwise directions respectively) and contains 64, 385 and 64 grid points in each of the corresponding dimensions. To collect mean data, the DNS was run for 100 eddy turnover times after the decay of transient startup effects, where here we define this timescale as $\delta_\Omega/u_\tau$, with $\delta_\Omega=0.025h$. The time-periodic mean velocity profile is obtained through averaging in the dimensions of spatial homogeneity (i.e.~the streamwise and spanwise directions), for each phase of wall motion. For resolvent analysis, the size of the numerical domain is $[0,h]\times[0,\tau_{tot}]$ with $h=1$ and $\tau_{tot}$ being the length of the time horizon. 

For performing both sparse and non-sparse space-time resolvent analysis, we consider wavelengths corresponding to the extent of the DNS computational domain, i.e.~$\lambda_x = 6\pi\delta_\Omega\ \approx 0.471 h$ and $\lambda_z = 3\pi\delta_\Omega\ \approx 0.236 h$. That is, we look at the largest three-dimensional structure that the DNS would be capable of resolving. The reason for focusing on such large structures in $x$ and $z$ is that we expect that they will correspond to space-time resolvent modes that also have a large extent in $t$ and $y$, allowing for the distinction between the sparse and non-sparse variants to be clearly observed.

\begin{figure}
\centering {
\vspace{-0.cm}
{\hspace*{-1.3cm}\includegraphics[width= 1.15\textwidth]{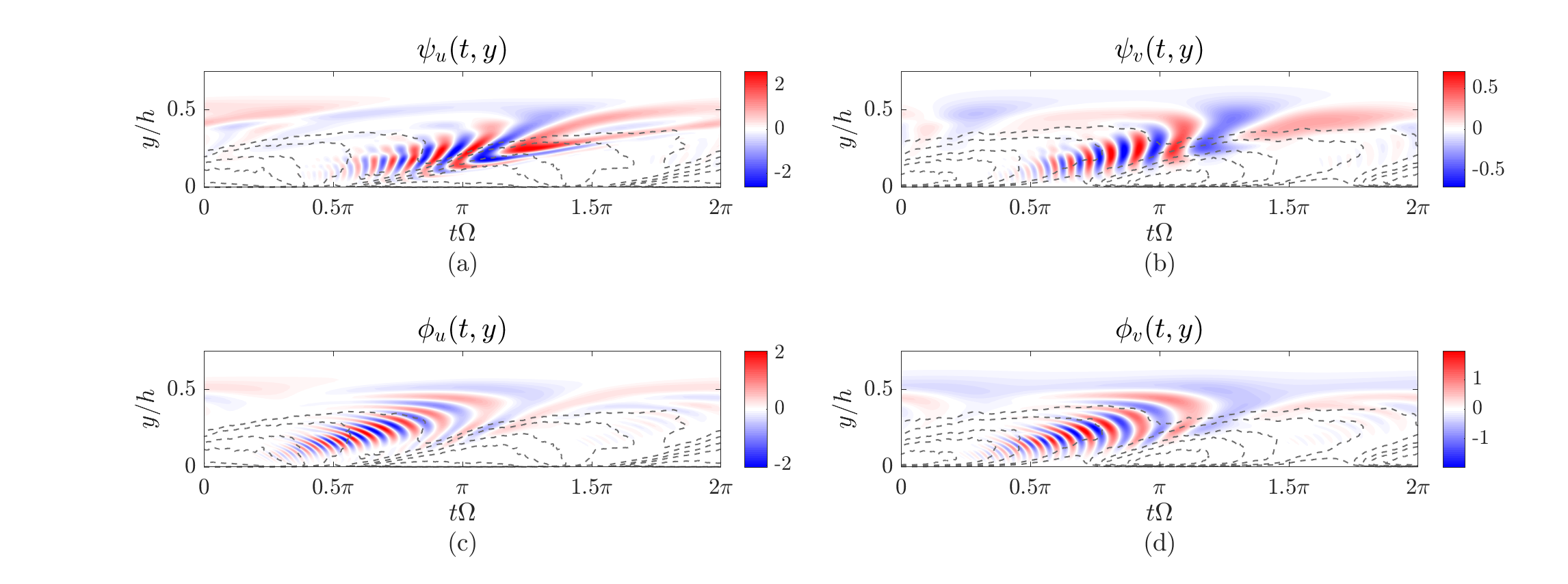}} }
\vspace{-0.8cm}
\caption{Real component of leading space-time resolvent (\textit{a,b}) response $\boldsymbol{\psi}_1$  and (\textit{c,d}) forcing $\boldsymbol{\phi}_1$  modes of the resolvent operator in a turbulent Stokes boundary layer with $\lambda_x/h$=0.471 and $\lambda_z/h$=0.236 during a full oscillating cycle. The grey dashed lines represent the (\textit{a,c}) streamwise and (\textit{b,d}) wall-normal  turbulence intensity, with contours at \{15\%,20\%30\%,40\%,60\%\} and \{22\%,30\%,45\%,55\%,70\%\}  of the maximum value of $u_{rms}$ and $v_{rms}$, respectively. 
}
\label{fig:standard_stokes}
\vspace{-0.4cm}
\end{figure}

For the non-sparse implementation of resolvent analysis, the domain is discretised using $N_y=105$ collocation points in the wall-normal axis and $N_t=651$ collocation points in the temporal dimension over a time window comprised of one oscillating cycle. 
The streamwise and wall-normal velocity components of the leading forcing and response modes obtained from this analysis are shown in figure~\ref{fig:standard_stokes}, along with the corresponding streamwise and wall-normal turbulence intensity computed from the DNS. Observe that the identified coherent structures do not correspond to a single temporal Fourier mode. Instead, they are concentrated over approximately half of the full oscillation period. Although not shown here, the second leading resolvent mode depicts a coherent structure of almost identical energy content that is similar to the leading modes in figure~\ref{fig:standard_stokes}, with a relative temporal shift of a half-period.

The location of the modes along the $y$-axis shifts away from the wall as the mean flow evolves over time. This angle in the $y-t$ plane appears to be similar to that observed for the turbulence intensities, suggesting that the mode depicts an energy transfer mechanism away from the wall that is present in the full nonlinear system. These modes are not concentrated in the near-wall region where the mean shear and turbulence intensity is largest. In this region, the turbulent energy content is likely dominated by  structures at smaller lengthscales than the wavelengths considered here. We do observe, however, in figure~\ref{fig:standard_stokes}(\textit{a)} that the streamwise velocity response is amplified as the mode enters a region of higher streamwise turbulence intensity at $(t\Omega,y/h)\approx (\pi,0.2)$. In addition, the disturbance stream functions identified in \cite{stabilityStokes2002} for a flat Stokes boundary layer are similar to the behavior of the space-time modes presented in figure~\ref{fig:standard_stokes} in two ways: first, the spatial shift of the structures along the wall-normal direction; second, the time-dependent oscillation frequency of these structures. The latter property will be addressed in more depth later in this section.

For the sparse implementation of space-time resolvent analysis at the same wavelengths, we adopt the same time horizon as in the non-sparse case, and a numerical domain with $N_y=101$ collocation points in the wall-normal axis and $N_t=631$ collocation points in the temporal dimension, with a sparsity ratio $\gamma=0.001$ ($\alpha = 0.625$). The location of maximum mode amplitude along the $y$-axis is indicated with a horizontal dotted line, and the value is indicated for each subplot with the symbol with $y_0/h$ and its equivalent in inner units $y_0^+$ for reference.

\begin{figure}
\centering {
\vspace{-0.cm}
{\hspace*{-1.cm}\includegraphics[width= 1.08\textwidth]{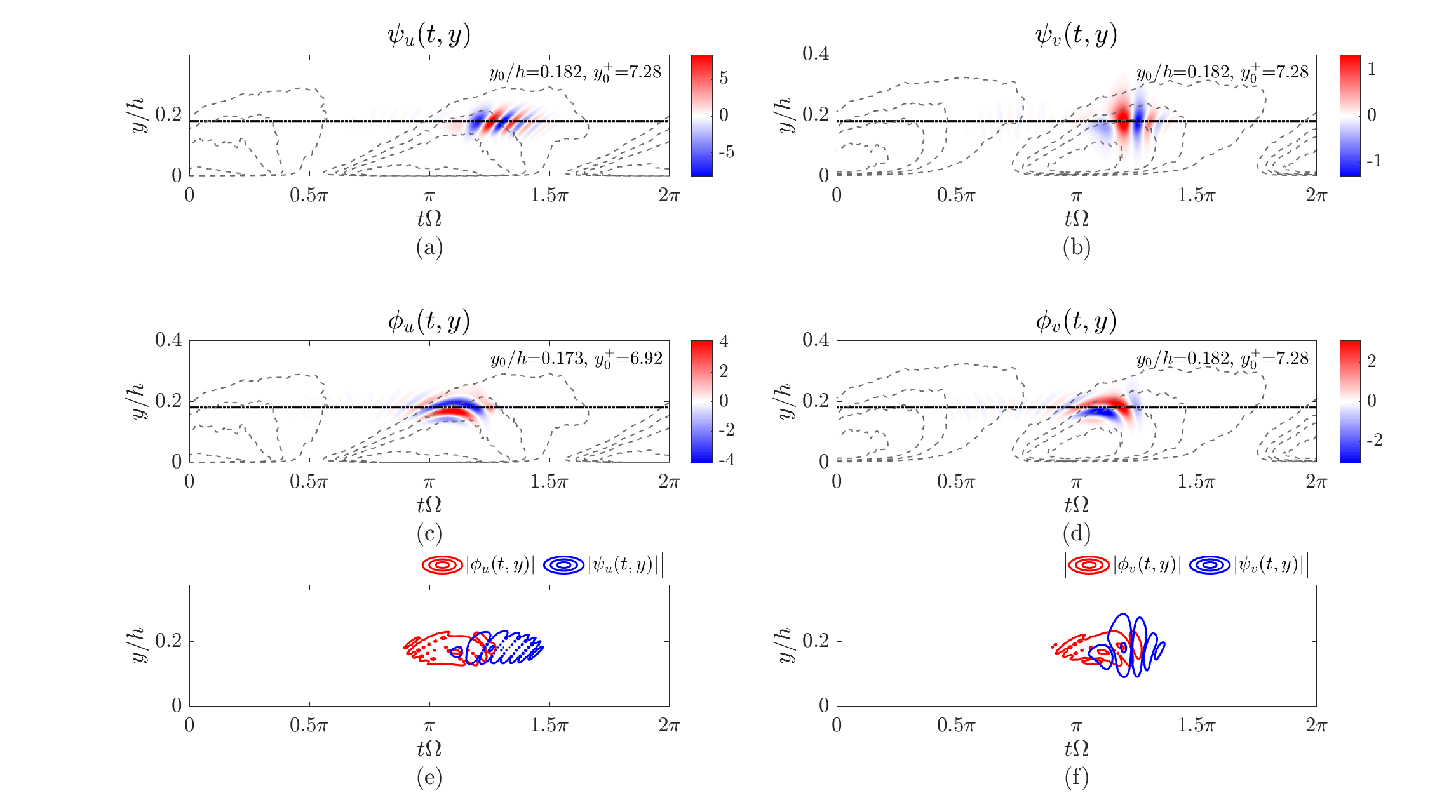}} }
\vspace{-0.4cm}
\caption{Real component of sparse space-time leading (\textit{a,b}) response $\boldsymbol{\psi}_1$  and (\textit{c,d}) forcing $\boldsymbol{\phi}_1$ of the resolvent operator for a turbulent Stokes boundary layer with $\lambda_x/h$=0.471 and $\lambda_z/h$=0.236, computed with a sparsity ratio $\gamma=0.001$, across a full oscillation cycle. Subplots (\textit{e,f}) show contour levels of the absolute value of leading forcing (red) and response (blue)  modes. The grey dashed lines represent the corresponding streamwise and wall-normal turbulence intensity, with contour levels at \{23\%,32\%40\%,50\%,70\%\} and \{35\%,45\%,60\%,75\%\} of the maximum value of $u_{rms}$ and $v_{rms}$, respectively. Horizontal dotted lines in (\textit{a-d}) indicate the $y$-location where each mode component achieves its maximum amplitude ($y_0/h$).
}
\vspace{-0.3cm}
\label{fig:sparse_stokes}
\end{figure}

The leading sparse resolvent modes are shown in figure~\ref{fig:sparse_stokes}, 
again overlaid with contour levels of the streamwise and wall-normal turbulence intensity. We observe that at these streamwise and spanwise wavelengths, the structure of the leading resolvent modes differs substantially between the 
non-sparse and sparse analyses. In figure~\ref{fig:sparse_stokes}, the sparse analysis identifies localised structures for which the streamwise and wall-normal components are concentrated  within one quarter of the total temporal domain, and within a small spatial region near $y/h=0.2$ (note that the vertical range shown in figure~\ref{fig:sparse_stokes} is smaller than in figure~\ref{fig:standard_stokes}).
Moreover, both forcing mode components precede their corresponding response modes (shown directly in figure~\ref{fig:sparse_stokes}(\textit{e-f})), with peak response in $v$ preceding that for $u$, which was also observed in the sparse modes computed for statistically-stationary channel flow in \S\ref{sec:resultsChannel2d}. 
Comparing the sparse modes with the non-sparse counterparts in figure~\ref{fig:standard_stokes}, the sparse analysis favours structures that are closer to the peak of the turbulence intensity. 

\begin{figure}
\centering {
\vspace{-0.cm}
{\hspace*{0cm}\includegraphics[width= 0.95\textwidth]{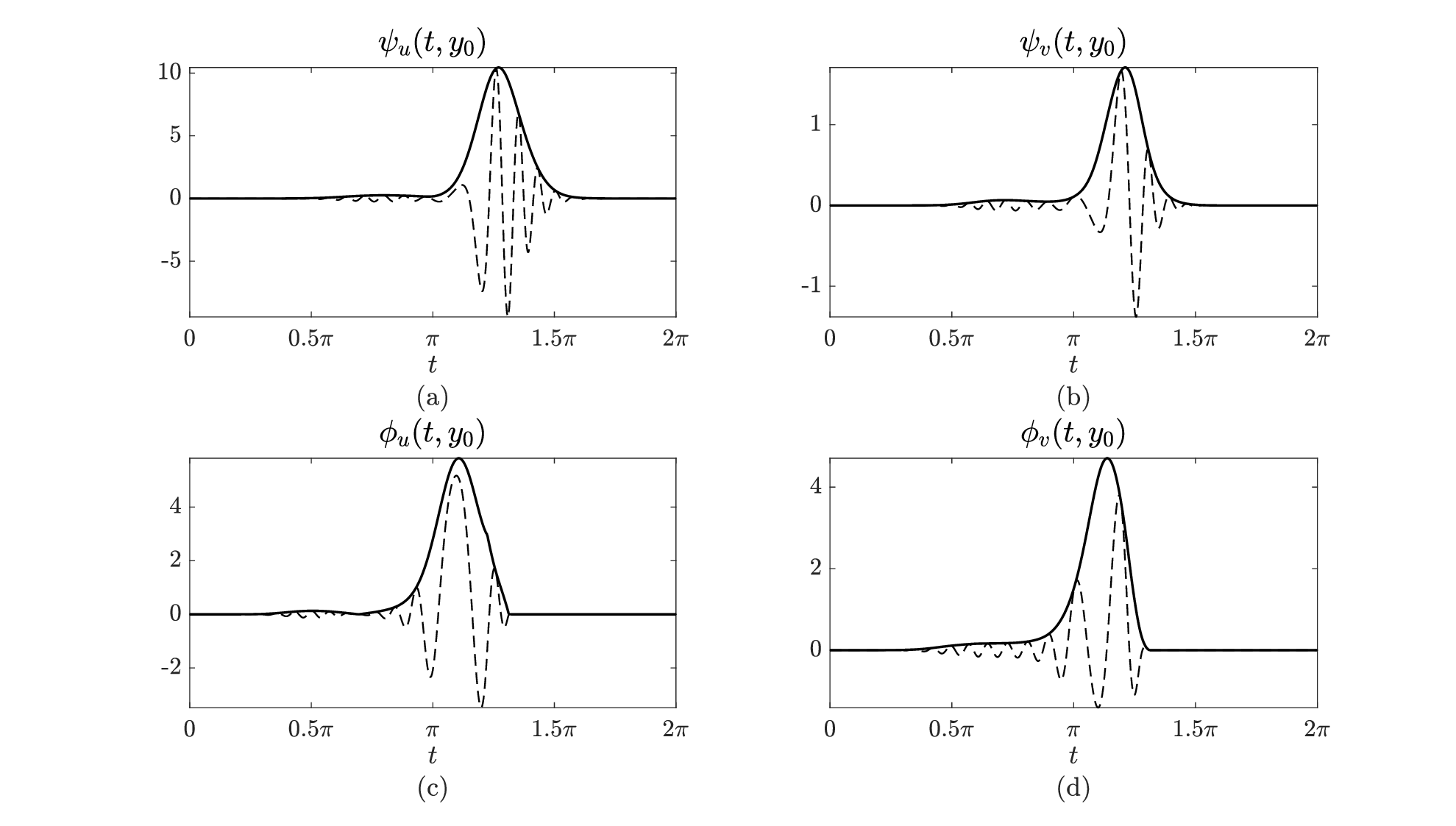}} }
\vspace{-0.1cm}
\caption{Absolute (solid) and real (dashed) components of the temporal evolution of the sparse modes shown in figure~\ref{fig:sparse_stokes}(\textit{a-d}), at the $y$-locations of maximum amplitude ($y_0$). 
}
\label{fig:sparse_stokes_cross}
\end{figure}

Figure~\ref{fig:sparse_stokes_cross} shows cross-section of these sparse modes along the temporal axis at the $y$-locations where they achieve their maximum amplitude (see the dotted lines indicated in figure~\ref{fig:sparse_stokes}). Comparing figures \ref{fig:spaceTime_turbchannel_sparse} and \ref{fig:sparse_stokes_cross}, we observe greater variation in the phase both within and across each mode component for the time-varying Stokes boundary layer configuration. For example, the streamwise velocity response in figure~\ref{fig:sparse_stokes_cross}(\textit{a}) shows that the rate of phase variation increases over time. The phase variation in the streamwise component of the sparse forcing mode is much slower, which figure~\ref{fig:sparse_stokes}(\textit{c}) shows is due to the almost horizontal inclination of this component. As was the case in figure~\ref{fig:spaceTime_turbchannel_sparse}, we observe that the forcing mode components have envelopes that are less symmetric in time, abruptly decaying to zero at a given cutoff time. 

\begin{figure}
\centering {
\vspace{0.2cm}
{\hspace*{-0.cm}\includegraphics[width= 1\textwidth]{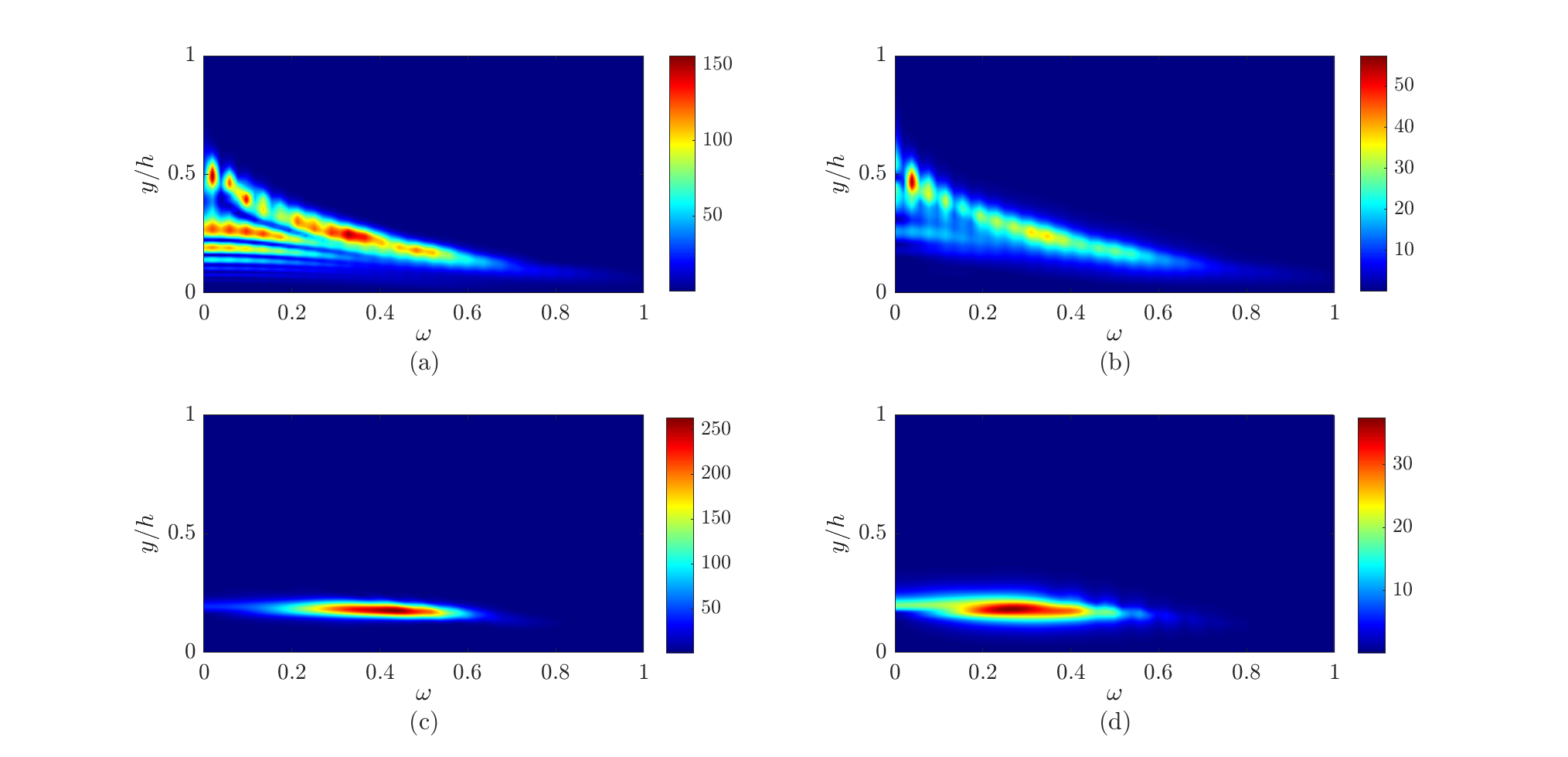}} }
\vspace{-0.4cm}
\caption{Spectral power of the streamwise ($u$) (a,c) and wall-normal ($v$) (b,d) components of the non-sparse (top row) and sparse (bottom row) spatiotemporal resolvent modes presented in figures~\ref{fig:standard_stokes} and \ref{fig:sparse_stokes}, respectively. A discrete Fourier transform is computed during one oscillating period at each of the locations of the $y$-axis in the numerical domain. 
}
\vspace{-0.cm}
\label{fig:fft_stokes}
\end{figure}

Figure~\ref{fig:fft_stokes} presents the spectral power of the non-sparse (top row) and sparse (bottom row) response modes shown in figures~\ref{fig:standard_stokes} and \ref{fig:sparse_stokes}, respectively. Each plot is generated via a temporal Fourier transform at each spatial location along the wall-normal axis ($y$). On the one hand, the regions of higher spectral power corresponding to the sparse modes in figure~\ref{fig:fft_stokes}(\textit{c,f}) are localized in $y$, but consist of a relatively broad frequency range. Conversely, the regions corresponding to the location of maximum spectral power of the non-sparse modes do not show the same degree of localization in $y$, and also exhibit several distinct peaks in the ($\omega, y/h$) plane. Furthermore, both the non-sparse and sparse spatiotemporal analyses identify continuous regions of active frequencies. Indeed, the identified modes in figures~\ref{fig:standard_stokes} and \ref{fig:sparse_stokes} depict structures with time-varying  frequencies. This evolution of the frequency within a single space-time mode would not be directly captured via harmonic resolvent analysis, where each component mode is associated with a single (discrete) temporal frequency (though triadic interactions between different frequencies could be isolated with this approach). Instead, the superposition of a considerable amount of temporal frequencies would be required to capture the dominant spatiotemporal modes identified by the direct space-time analysis. This suggests that the present approach could be more insightful when applied to flows that do not necessarily possess sparse frequency content, such as the configuration considered in this section. 

\subsection{Space-time resolvent analysis of a turbulent channel flow with sudden lateral pressure gradient}
\label{sec:results3Dchannel}
\begin{figure}
\centering {
\vspace{0.3cm}
{\hspace*{0cm}\includegraphics[width= 1\textwidth]{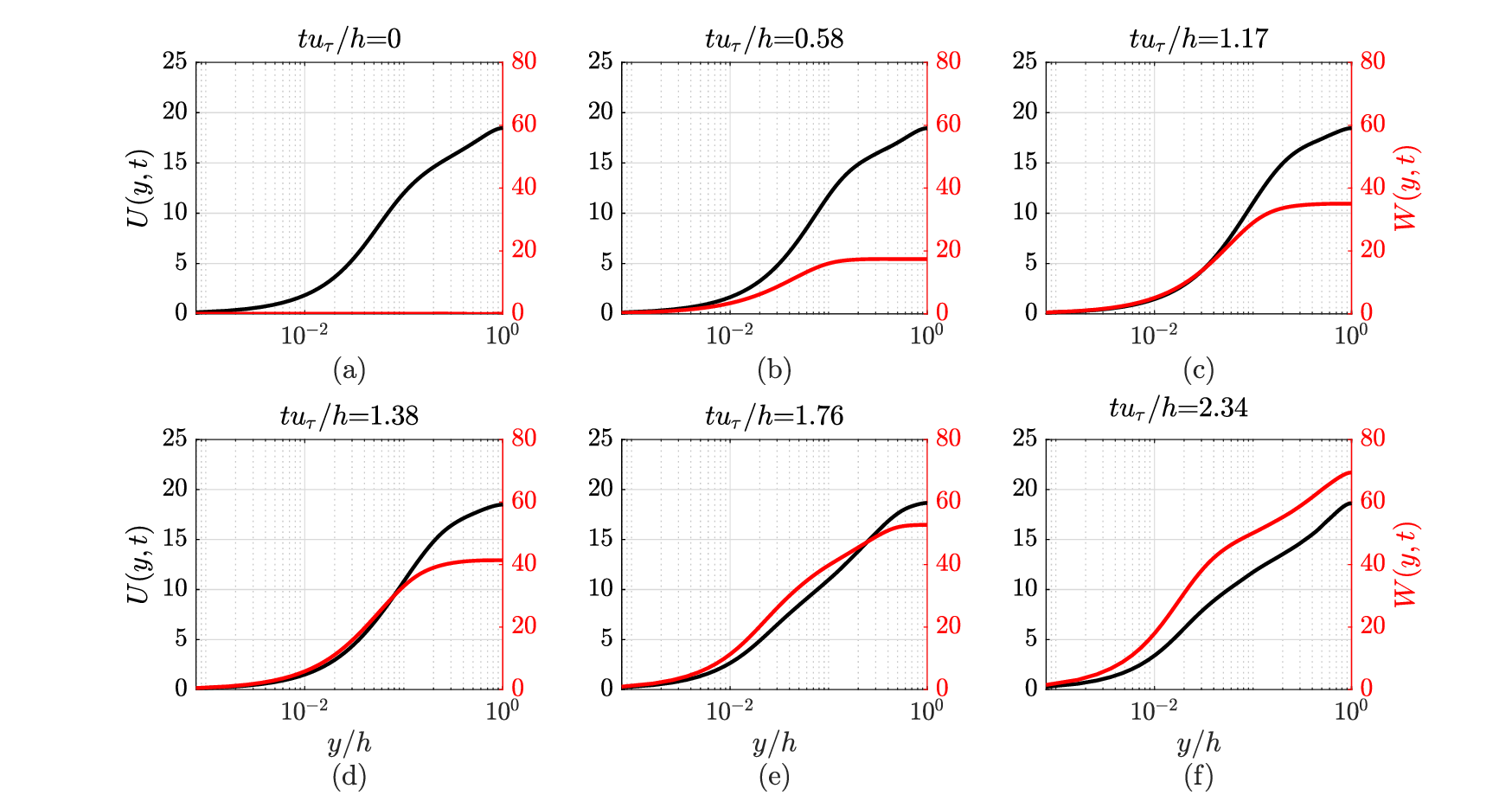}} }
\vspace{-0.2cm}
\caption{Temporal evolution of the streamwise $U(t,y)$ (black) and spanwise $W(t,y)$ (red) mean velocity profiles obtained by DNS of a turbulent channel flow at $\Rey_\tau=186$ with sudden lateral pressure gradient at $t=0$ with $\pp p/\pp z=\Pi (\pp p/\pp x)$ and $\Pi=30$ at $tu_\tau/h\in\lbrace 0,0.58,1.17,1.38,1.76,2.34\rbrace$ \citep{lozanoduran2021boundary3d}.
}
\label{fig:3dmean}
\end{figure}

The last case considered in this work 
is a 
fully-developed turbulent channel flow at $\Rey_\tau=186$ that is subjected to a sudden lateral pressure gradient at $t=0$. The spanwise pressure gradient is related to the fixed streamwise gradient by $\pp p/\pp z=\Pi (\pp p/\pp x)$, where here we use $\Pi=30$. Additional information about this configuration can be found in \citet{moin1990boundary3d,lozanoduran2021boundary3d}. The turbulent mean flow therefore contains both streamwise $U(y,t)$ and spanwise $W(y,t)$ nonzero components, which evolve over time until the system reaches its new statistically-stationary state. These velocity components are shown in figure~\ref{fig:3dmean}. 

%
To form the resolvent operator, the temporal domain is implicitly nondimensionalised by the the initial friction velocity $u_{\tau,0}$ at $t=0$, and $h$, and it is determined in two steps to achieve numerical accuracy. It is first defined as the interval $[-0.58,2.34]h/u_\tau$ to extend the numerical domain before the initial condition to include time prior to the application of the spanwise pressure gradient. Thus, for $t<0$ the base flow is constant, with $U(t<0,y)=U(T=0,y)$ and $W(t<0,y)=0$, and $\pp p/\pp z(t<0)=0$. After a first analysis is conducted, the time domain is restricted to a smaller temporal window near the  observed temporal location of the leading modes that is large enough to produce unaltered results with a larger temporal resolution. Note that here we implement a explicit Euler finite-differentiation scheme in the temporal dimension. Thus, the final numerical domain used in this analysis is given by $[0,h]\times [0.4,2]h/u_\tau$. 

As was the case for statistically-stationary channel flow studied in \S\ref{sec:resultsChannel1d}-\ref{sec:resultsChannel2d}, we concentrate our analysis on streamwise and spanwise wavenumbers corresponding to the typical size of near-wall streaks. After the pressure gradient is applied, the magnitude and direction of the mean flow changes, meaning that the wavelengths associated with near-wall streaks is also expected to change. Assuming that near-wall streaks adjust their size and orientation with the mean, we can select wavelengths corresponding to the typical size of near-wall streaks at a certain point in time \citep{ballouz2024wavelet} (though as before, we are not guaranteed that the wall-normal location of the resulting modes will match those typical of such streaks). Here, we select $\lambda^+_x=189$ and $\lambda^+_z=1890$, which correspond to the typical length of the near-wall streaks at a time instance $t u_\tau/h=1.38$. 

\begin{figure}
\centering {
\vspace{-0cm}
{\hspace*{-1.3cm}\includegraphics[width= 1.08\textwidth]{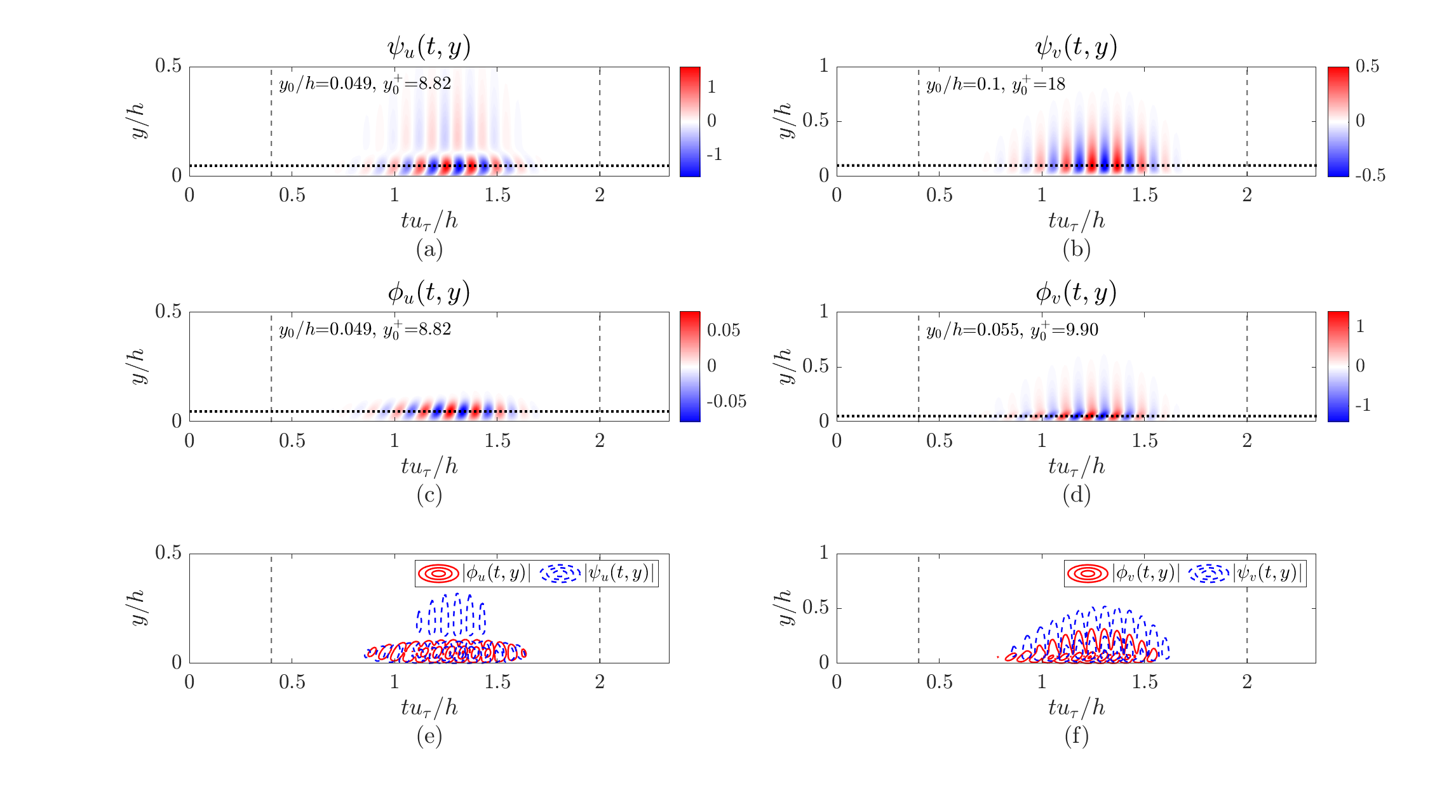}} }
\vspace{-0.85cm}
\caption{Real component of space-time $u$ and $v$ velocity components of the leading (\textit{a,b}) response $\boldsymbol{\psi}_1$  and (\textit{c,d}) forcing $\boldsymbol{\phi}_1$ modes of the space-time resolvent operator for a channel flow with sudden lateral pressure gradient with $\Rey_\tau=186$, and $\lambda^+_x=189$, $\lambda^+_z =1890$. (\textit{e,f}) Contour levels of the absolute value of the leading forcing (red) and response (blue) modes. Horizontal dotted lines in (\textit{a-d}) indicate the $y$-location where each mode component achieves its maximum amplitude ($y_0/h$). The grey dashed lines represent the boundaries of the temporal numerical domain.
}
\label{fig:spaceTime_3dchannel}
\vspace{-0.cm}
\end{figure}

For the non-sparse space-time resolvent analysis, the domain is discretised using $N_y=132$ and $N_t=500$ collocation points. The location of maximum mode amplitude along the $y$-axis is indicated with a horizontal dotted line, and the value is indicated for each subplot with the symbol with $y_0/h$ and its equivalent in inner units $y_0^+$ for reference. Leading resolvent mode components for this analysis are shown in figure~\ref{fig:spaceTime_3dchannel}. We observe that the modes are localised in time, centered near to the the nominal time ($t u_\tau/h=1.38$) where the streamwise and spanwise lengthscales of the modes coincides with near-wall streaks. The structure of the modes shares some similarities with those observed for both the non-sparse and sparse analyses of stationary channel flow (figures~\ref{fig:spaceTime_turbchannel} and \ref{fig:spaceTime_turbchannel_sparse}). In all cases, the wall-normal response consists of vertically-aligned structures, while the forcing is inclined upstream in the $y-t$ plane. The $u$-component of the response appears to incline further backwards as time progresses in both figures~\ref{fig:spaceTime_turbchannel_sparse}(\textit{a}) and \ref{fig:spaceTime_3dchannel}(\textit{a}), though the latter does appear to move towards and move away from the wall as observed in the former.  Note that the $u$-component plotted in figure~\ref{fig:spaceTime_3dchannel} no longer entirely corresponds to the streamwise direction, due to the spanwise pressure gradient. 

Figure~\ref{fig:spaceTime_3dchannel}(\textit{e-f}) show that the forcing and response mode components are all located  over approximately the same time interval for this case. This is different from the sparse analysis of statistically-stationary channel flow (see figure~\ref{fig:spaceTime_turbchannel_sparse}(\textit{e-f})), where the forcing mode components decayed before their responses did, and the wall-normal forcing and response components started before the streamwise components. These differences can likely be attributed to the differences in how localised modes are obtained; for the 3D channel configuration we are not explicitly promoting sparse modes, but rather the modes are aligning with a region of time corresponding to a mean flow with maximum linear energy amplification for structures at the specified spatial scales. This can be further explored by applying the sparse variant to this configuration.

\begin{figure}
\centering {
\vspace{0.2cm}
{\hspace*{-1.1cm}\includegraphics[width= 1.02\textwidth]{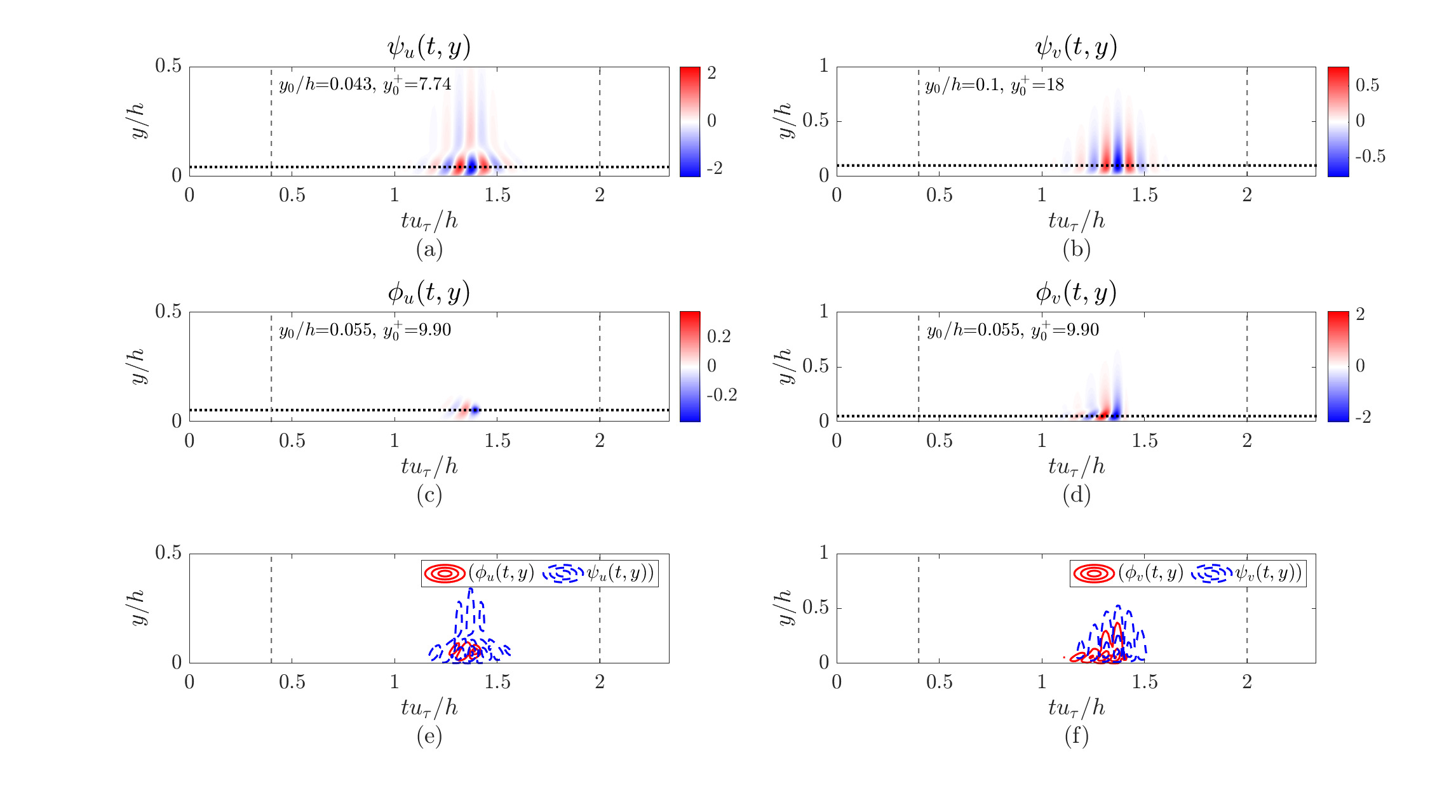}} }
\vspace{-0.4cm}
\caption{Real component of sparse space-time $u$ and $v$ velocity components of the leading (\textit{a,b}) response $\boldsymbol{\psi}_1$ and (\textit{c,d}) forcing $\boldsymbol{\phi}_1$  modes of the space-time resolvent for a channel flow with sudden lateral pressure gradient with $\Rey_\tau=186$, and $\lambda^+_x=189$, $\lambda^+_z =1890$ with a sparsity ratio $\gamma=0.001$ ($\alpha = 0.781$). (\textit{e,f}) Contour levels of absolute value of the leading forcing (red) and response (blue) modes. Horizontal dotted lines in (\textit{a-d}) indicate the $y$-location where each mode component achieves its maximum amplitude ($y_0/h$). The grey dashed lines represent the boundaries of the temporal numerical domain.
}
\vspace{-0.cm}
\label{fig:spaceTime_3dchannel_sparse}
\end{figure}

For the sparse implementation of space-time resolvent analysis, we adopt the same wavelengths and a sparsity ratio $\gamma=0.001$ ($\alpha = 0.781$). The domain is discretised using $N_y=130$ and $N_t=500$ collocation points. The leading sparse modes are shown in figure~\ref{fig:spaceTime_3dchannel_sparse}, where both the $u$ and $v$ components  display a similar structure, but with more temporal localisation, to those shown for the non-sparse analysis in figure~\ref{fig:spaceTime_3dchannel}. The location of maximum mode amplitude along the $y$-axis is indicated with a horizontal dotted line, and the value is indicated for each subplot with the symbol with $y_0/h$ and its equivalent in inner units $y_0^+$ for reference. 

The sparse analysis thus appears to identify the same mechanism, localised around the same region in  time. 
Unlike the non-sparse version, here we observe in figure~\ref{fig:spaceTime_3dchannel_sparse}(\textit{e-f}) differences in the temporal footprints of each mode component, with  the decay in amplitude of the both components of the forcing preceding the decay of the response.

\begin{figure}
\centering {
\vspace{-0.cm}
{\hspace*{0cm}\includegraphics[width= 1\textwidth]{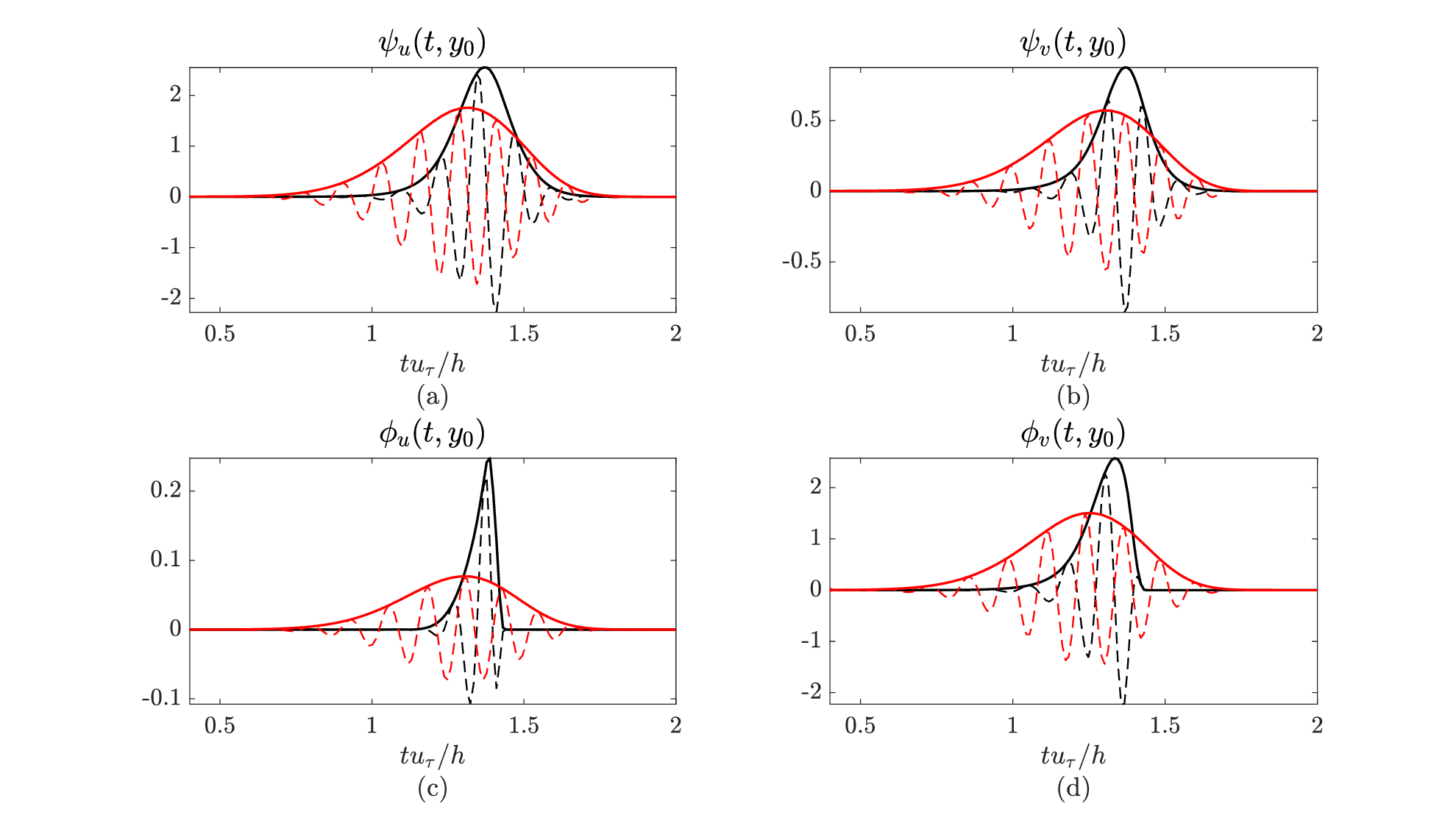}} }
\vspace{-0.4cm}
\caption{Absolute (solid) and real (dashed) components of the temporal evolution of the non-sparse modes shown in figure~\ref{fig:spaceTime_3dchannel}(\textit{a-d}) (red) and sparse modes shown in figure~\ref{fig:spaceTime_3dchannel_sparse}(\textit{a-d}) (black), at the $y$-locations of maximum amplitude.
}
\label{fig:spaceTime_3dchannel_sparse_cross}
\end{figure}

The cross-sections along the time axis of the modes contained in figures~\ref{fig:spaceTime_3dchannel} and \ref{fig:spaceTime_3dchannel_sparse} are shown in figure~\ref{fig:spaceTime_3dchannel_sparse_cross} in red and black, respectively. The amplitudes of both the non-sparse and sparse modes adopt qualitatively similar envelopes, with the sparse variant being narrower, and centered at a slightly later time. 
We emphasise that unlike the profiles shown in figure~\ref{fig:spaceTime_turbchannel_sparse_cross}, here the time-location of these modes is not arbitrary, and corresponds to a region of time where the mean profile enables the largest linear amplification. The phase variation is similar across all mode components, indicating a characteristic frequency maximising amplification. Here, we thus find that both sparse and non-sparse space-time resolvent appear to identify the same amplification mechanism, with the addition of the $l_1$-norm term in the objective function restricting the sparse modes to a smaller temporal window. 

\begin{figure}
\centering {
\vspace{-0.cm}
{\hspace*{0cm}\includegraphics[width= 0.6\textwidth]{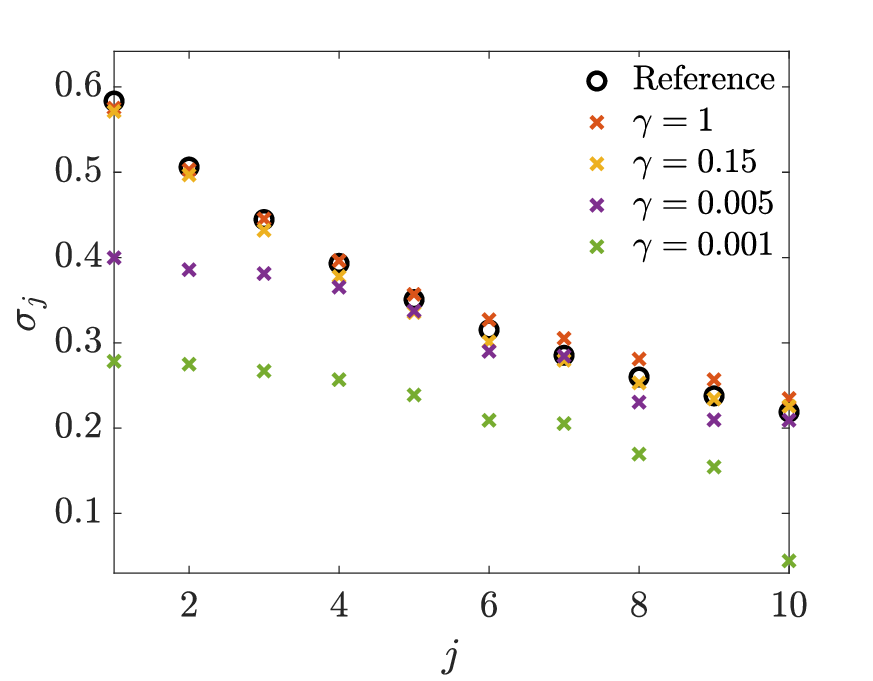}} }
\vspace{-0.1cm}
\caption{Leading $10$ singular values of the non-sparse and sparse (with $\gamma \in \lbrace 0.001,0.15 \rbrace$) space-time resolvent operator for a channel flow with sudden lateral pressure gradient with $\Rey_\tau=186$, and $\lambda^+_x=189$, $\lambda^+_z =1890$. The reference values correspond to the analysis described in \citet{ballouz2024wavelet}.
}
\label{fig:spaceTime_3dchannel_singVals}
\end{figure}

Lastly, to demonstrate the numerical convergence of the results presented in this section, figure~\ref{fig:spaceTime_3dchannel_singVals} presents the first ten singular values of the non-sparse and sparse with $\gamma \in \lbrace 0.001,0.15 \rbrace$  space-time resolvent operators against the leading singular values obtained in the study described in \citet{ballouz2024wavelet}, where the numerical methods utilize a wavelet transform in time. The corresponding leading modes coincide with figure~\ref{fig:spaceTime_3dchannel} and figure~\ref{fig:spaceTime_3dchannel_sparse} in the non-sparse and sparse with $\gamma = 0.001$ ($\alpha = 0.781$) implementation of the space-time analysis, respectively. First, observe that the singular values of the non-sparse analysis closely match the reference values. Second, a sparsity ratio of $\gamma=0.15$ ($\alpha = 0.004$) is sufficient large to obtain similar amplification levels to these reference values (note that the corresponding modes are not shown in this paper). This agreement also confirms the convergence of the sparse analysis. However, the singular values retrieved from the sparse analysis with $\gamma = 0.001$ ($\alpha = 0.781$) and $\gamma = 0.005$ ($\alpha = 0.258$), retrieve only a portion of the reference values. In particular, the leading reference value is larger than the leading sparse singular value with $\gamma = 0.001$ by a factor of 2.0958, and we observe that this ratio decreases for higher-order singular values. The fact that the sparse and non-sparse analyses yield similar results for this case is likely due to the highly transient nature of the flow, where the leading resolvent modes are naturally localised even without explicitly promoting sparsity.

\section{Discussion and Conclusions}
\label{sec:conclusions}
In this work, we have proposed a sparse space-time variant of resolvent analysis that can 
identify time-localised coherent structures. These spatio-temporal structures correspond to the inputs and outputs that optimise an objective function that promotes large linear energy amplification, while also promoting the localisation in space and time of these structures.
Localisation is achieved through the addition of a  sparsity-promoting $l_1$-norm term to the standard optimisation problem used for resolvent-type analyses. The new optimisation problem takes the form of a nonlinear eigenproblem, for which the optimal solution is achieved through an inverse power method. 

We have demonstrated  the implementation of this sparse variant of resolvent analysis on several different configurations, demonstrating that we obtain sparse modes first in the spatial domain, and also in the spatial and temporal domains when using a generalised space-time formulation of resolvent analysis. 
The first case studied was a statistically-stationary  turbulent channel flow, a canonical configuration that has received substantial prior study in the context of resolvent analysis (and for many other forms of analysis). When using the standard Fourier transform in time, the proposed sparse resolvent analysis identified spatially-sparse modes, in either the wall-normal or  wall-normal and spanwise dimensions, depending on the problem setup. This provides a means for identifying localised spatial structures corresponding to similar amplification as the non-localised structures identified in standard resolvent analysis. For the one-dimensional channel flow analysis, this sparsity simply resulted in modes concentrated about one of the two critical layers. In this case, the symmetry of the configuration means that leading resolvent modes come with a multiplicity of two, and sparse resolvent analysis found a basis for this subspace where each basis vector was as sparse as possible. For the two-dimensional analysis where the spanwise direction was explicitly discretised, sparse resolvent analysis found a leading resolvent mode that consisted of a primary streamwise vortex, surrounded by a pair of streamwise streaks, with comparable total amplification to the Fourier mode identified from standard resolvent analysis. This suggests that the ordered, spanwise-repetition of such structures are not essential for their appearance in turbulent flows (under the assumption that linear mechanisms are at least partially responsible for their emergence). Indeed, this is consistent with analysis from turbulent flow data, where spanwise correlations are observed to decay. Indeed, correlations at a wall-normal location corresponding to near-wall streaks are observed to decay \citep{kim1987turbulence} over a spanwise length scale similar to the streak width itself ($\lambda_z^+ \approx 100$), which is consistent with the spanwise localisation observed for the leading resolvent response mode in figure \ref{fig:spanwiseChann}(\textit{b}).

Next, a space-time generalisation of resolvent analysis was applied to the same channel flow configuration. Being a statistically-stationary system, each of the space-time modes converged to a single temporal Fourier mode in the absence of sparsity-promotion. Therefore, the space-time resolvent modes are equivalent to the standard resolvent modes compiled across all frequencies permissible by the prescribed time horizon. The sparsity-promoting variant of this space-time resolvent analysis yielded structures that displayed localisation in both the spatial and temporal dimensions. 
These modes again reflect the fact that linear energy amplification can arise from forcing that is limited to a localised region in time. These time-localised structures  exhibit features that cannot be seen in Fourier modes, such as the time-evolution of inclination angles and wall-normal locations of structures within resolvent modes. Note that the change in the inclination angles of the modes can also be interpreted as regions of the modes with different wall-normal locations evolving with different wavespeeds, consistent with the distribution of the mean velocity across the wall-normal extent of the modes. This phenomena is again not observable in standard Fourier-based resolvent analysis. It was noted in \S\ref{sec:resultsChannel2d} that when streamwise and spanwise wavelengths are chosen to be typical of near-wall streaks, the resulting resolvent modes are centered much further from the wall than where such structures are typically found. However, the corresponding sparse space-time mode components exhibited peak amplitudes much closer to the wall, with the streamwise component of the leading response mode having a peak at $y^+\approx 17.5$, which is more consistent with the location of near-wall streaks. Future work could explore in more detail the quantitative  extent to which sparse modes predict phenomena present in turbulent channel flow data.

Beyond statistically-stationary systems, we next explored the application of our proposed methodology to systems with a time-varying mean. 
The first such system that was considered was a time-periodic turbulent Stokes boundary layer. 
In this case, the non-sparse and sparse variants identified leading resolvent modes that were qualitatively different, 
with the sparse models being significantly more localised in both space and time.  Our analysis focused on relatively large-scale structures by specifying streamwise and spanwise wavelengths corresponding to the size of the domain used for DNS. While not shown here, we found that this difference is was less apparent for smaller wavelengths, where even the non-sparse variant identified structures that were localised in time, such as the case considered in \citet{ballouz2024wavelet}. 
For the case  considered here, the sparse mode components identified were also localised near the peak in turbulence intensity at the wall-normal location where they were situated, suggesting that they identify a mechanism relevant to the generation and amplification of turbulent features at this temporal phase of the system. 

The final set of results considered a non-periodic,  time-evolving turbulent channel flow subjected to a sudden lateral pressure gradient. This flow was studied during a temporal window in which the flow transitions between two statistically-stationary states. It was found that both  the non-sparse and sparse modes showed temporal localisation, presumably due to the choice of wavelengths permitting largest energy amplification in a certain time window corresponding to a given orientation of the mean. 
Overall, we have demonstrated that spatio-temporally sparse resolvent analysis can identify features that can either be qualitatively similar or distinct from the non-sparse equivalent. 

Across all of the examples considered, there are several common observations that can be made. First, across all configurations, the streamwise velocity component of the response modes are larger than the wall-normal component. Furthermore, for the time-localized modes, the peak in the amplitude of the streamwise velocity typically occurrs slightly after the peak in the wall-normal component. This is consistent with the lift-up mechanism that transfers momentum between these components for cases with nonzero spanwise wavenumbers. For cases where sparsity is promoted in time, we observe that response mode shapes are typically localized within temporal envelopes that are approximately Gaussian in shape, with modes exhibiting several periods of oscillation within such envelopes. For all but the Stokes boundary layer case, the frequency of these oscillations remains approximately constant in time throughout the mode evolution. For the Stokes boundary layer, the sparse response mode oscillation frequency is found to increase continuously with time, approximately doubling throughout the time-evolution of the mode. This suggests that for systems with large, rapid variations in the mean velocity field, analysis of linear amplification mechanisms could be more difficult to capture in the frequency (rather than time) domain, given the observed continual transfer of energy across frequencies.

To conclude, the contributions provided by both of the frameworks introduced in this manuscript, in terms of improved physical insight, deserve a separate discussion. Firstly, the main advantage of the space-time analysis with respect to traditional resolvent analysis resides in its applicability to not-only statistically-stationary systems, but also time-varying mean flows of any nature. This aspect enables the identification and characterisation of the relevant spatiotemporal coherent structures that are most highly amplified by the linearised governing equations. Secondly, the analysis also recognises the physically-relevant temporal instance at which coherent structures of a targeted lengthscale naturally arises. This was observed in the cases presented in \S\ref{sec:resultsStokesBL} and \S\ref{sec:results3Dchannel}. Thirdly, considering the application of the space-time framework to time-periodic systems, its capabilities surpass those of harmonic resolvent analysis, as it is able to infer triadic interactions between modes in which each of them can be associated with several temporal frequencies (see the cases presented in \S\ref{sec:resultsStokesBL} for a turbulent Stokes boundary layer). Lastly, even the application of the space-time analysis provides additional physical insight for time-invariant mean flows, as it highlights not only the coherent structures of maximal amplification, but also the associated temporal frequency within a prescribed temporal period (see figure~\ref{fig:spaceTime_turbchannel_sparse} for a turbulent channel flow). 

In addition, the incorporation of the sparsity-promoting term to the space-time analysis further expands its potential in a number of ways. For instance, by promoting localisation, it is possible to break down a given spatiotemporal coherent structure formed by an aggregate of temporal frequencies into smaller units (see case presented in figure~\ref{fig:sparse_stokes} for a turbulent Stokes boundary layer). This presumably allows for the study of complex linear energy-amplification mechanisms through the examination and characterisation of localised coherent structures of a reduced frequency content. Moreover, the exploitation of the sparsity-promoting space-time analysis could assist in the design of control strategies of targeted length- and temporal scales. Note that this analysis not only provides a localised structure that is optimal in terms of the energy gain (this was investigated in more depth in figures~\ref{fig:modes_restricted}-\ref{fig:singVals_restricted} in the study of the turbulent channel flow with discretisation in the spanwise dimension), but will also be identified in a physically-relevant temporal instance. Thus, this technique could be advantageous in the determination of the most optimal location for actuator and/or sensor placement. 

The focus of this work has been the development and demonstration of this method across a range of flow configurations. Additional investigations into each of these (and other) configurations are required to quantify the extent to which these methods can be used to predict and understand structures and statistics of such turbulent flows, over a wider range of spatial scales. There are several possible approaches towards assessing the capacity of sparse-resolvent-based models in predicting turbulent flow features. One approach involves a direct comparison between predicted mode structures from sparse resolvent analysis with a data-driven equivalent. This could involve modifying the optimisation problem associated with standard (space-only or spacetime) POD methods to include a sparsity-promoting term, or through comparison with previously-developed conditional \citep{schmidt2019conditional} or windowed \citep{frame2022space} space-time POD. Beyond this, sparse resolvent modes could also be used to develop simplified `turbulence kernels,' which may facilitate modeling similar structures to those captures using sets of standard resolvent modes \citep{sharma2013resolvent,luhar2014pressure,mckeon2017engine}, but while also capturing the decay in two-point correlations with increasing spatio-temporal separation that are typically observed in turbulent flows, a property  not readily captured with a small number of Fourier modes. Additionally, bases of sparse resolvent modes could also be utilized for real-time flow estimation tasks \citep{arun2023towards}, which may be particularly useful if needing to estimate localised features within large domains. The numerical methods employed in the present work have required the explicit formation and decomposition of linear operators discretised in both space and time. Future work could also investigate timestepping approaches to perform such analyses. 
Such methods, which have been applied for resolvent \citep{martini2020,farghadan2023scalable} and other linear analyses \citep{barkley2008optimal}, would likely be particularly computationally-advantageous for our methodology.

\section*{Acknowledgements}
This work was supported by the Air Force Office of Scientific Research grant FA9550-22-1-0109 and National Science Foundation Award number 2238770. The authors gratefully acknowledge the advanced computing resources made available through ACCESS project MCH230005.

\section*{Declaration of Interests}
The authors report no conflict of interest.

\appendix
\section{Notes on the inverse power method for nonlinear eigenproblem used in sparse resolvent analysis}\label{appA}
This appendix contains a detailed description of the inverse power method proposed by \citet{hein2010inverse} that is employed to solve the constrained optimisation problem posed by the sparse PCA algorithm, adapted to perform sparse resolvent analysis in \S\ref{sec:sparseResolvent}. We start by considering a general nonlinear eigenproblem of the form
\begin{equation}
    \label{eq:nonlinearEigen}
    r(\bsf)-\lambda s(\bsf) =0,
\end{equation}
where $\bsf$ represents the eigenfunction associated to the eigenvalue $\lambda$, and $r$ and $s$ are nonlinear operators. In order for $\bsf$ to be a solution to the nonlinear eigenproblem, $\bsf$ has to be a critical point of a functional, $F$. In this problem, the functional $F$ is expressed as
\begin{equation}
    \label{eq:functional}
    F(\bsf)=\frac{R(\bsf)}{S(\bsf)},
\end{equation}
where both $S$ and $R$ represent convex, non-negative, Lipschitz continuous functionals that satisfy the positive homogeneity property $R(\gamma \bsf) = \gamma R(\bsf)$ for $\gamma \geq 0$. 
 The critical points of $F(\bsf)$ are found at the locations $\bsf^c$ in which $\nabla F(\bsf^c)=0$. We can write this condition in terms of $R$ and $S$ as
\begin{equation}
\label{eq:criticalP}
    \nabla F(\bsf^c) = \frac{S(\bsf^c)\nabla R(\bsf^c)-R(\bsf^c)\nabla S(\bsf^c)}{S^2(\bsf^c)}=0.
\end{equation}
Rearranging the terms in \eqref{eq:criticalP} yields a form that assimilates the eigenproblem in \eqref{eq:nonlinearEigen} as 
\begin{equation}
\label{eq:nonlinearFunc}
    \nabla R(\bsf^c)-\frac{R(\bsf^c)}{S(\bsf^c)} \nabla S(\bsf^c)=0,   
\end{equation}
where we have adopted $r(\bsf^c)=\nabla R(\bsf^c)$, $s(\bsf^c)=\nabla S(\bsf^c)$ and $\lambda=R(\bsf^c)/S(\bsf^c)$. Note that in order for this definition to hold, $S$ must be continuously differentiable. In addition, in the case where $R(\bsf)$ and $S(\bsf)$ represent second-order functionals, \eqref{eq:nonlinearFunc} becomes a linear eigenproblem.

The nonlinear eigenproblem in \eqref{eq:nonlinearFunc} is solved using an inverse iteration, also referred to as inverse power method. For a linear operator $A$, this scheme takes the form A$\bsf^{k+1}=\bsf^k$, which
is equivalent to the solution to the optimisation problem 
\begin{equation}
\label{eq:IPM0}
    \bsf^{k+1}=\argmin_{\|\bsf\|_2\leq1} \left(\frac{1}{2}\langle\bsf,A\bsu\rangle-\langle \bsf,\bsf^k \rangle \right).
\end{equation}
 Considering the general eigenproblem presented in \eqref{eq:nonlinearEigen}, the iterative scheme utilised in this work is written as
\begin{equation}
\label{eq:IPM}
    \bsf^{k+1}=\argmin_{\|\bsf\|_2\leq1}\left(R(\bsf)-\langle \bsf,s(\bsf^k) \rangle \right). 
\end{equation}
According to the form of the eigenproblem shown in \eqref{eq:nonlinearFunc} the corresponding eigenvalue is computed as $\lambda^{k+1}=R(\bsf^{k+1})/S(\bsf^{k+1})$. Note that in our problem, the function $F(\bsf)$ takes the form 
\begin{equation}
\label{eq:functMin}
    F(\bsf)=\frac{(1-\alpha)\|\bsf\|_2+\alpha\|\bsf\|_1}{\|\mathcal{H}^* \bsf\|_2},
\end{equation}
for which the optimisation problem in \eqref{eq:IPM} is rewritten as the following convex optimization problem
\begin{equation}
\label{eq:IPMMin}
    \bsg^{k+1}=\argmin_{\|\bsf\|_2\leq1} \left[(1-\alpha)\|\bsf\|_2+\alpha\|\bsf\|_1-\lambda^k\langle \bsf,\bmu^k \rangle \right], 
\end{equation}
where
\begin{equation}
    \bmu^k=\frac{\Sigma \bsf^k}{\sqrt{\langle \bsf^k,\Sigma \bsf^k\rangle}},
\end{equation}
where for our problem, $\Sigma = \mathcal{H}\mathcal{H}^*$. 
This optimisation problem has the closed solution
\begin{equation}
    \label{eq:closedSol}
    \bsg_i^{k+1}=\frac{\text{sign}(\bmu_i^k)(\lambda^k|\bmu_i^k|-\alpha)_+}{s},
\end{equation}
with $s=\sqrt{\sum_{i=1}^n (\lambda^k|\bmu_i^k|-\alpha)^2_+}$ and $a_+=\max(0,a)$. Notice that $s$ now represents a scaling factor and is excluded from the method for simplicity. The algorithm that computes the optimal solution to the nonlinear eigenproblem in \eqref{eq:nonlinearFunc} using an inverse power method with the closed solution indicated in \eqref{eq:closedSol} for sparse resolvent analysis is presented in algorithm~\ref{alg:inverse}.

\begin{algorithm}
    \caption{Inverse power method for sparse resolvent analysis}\label{alg:inverse}
    \begin{algorithmic}[1]
        \vspace{0.1cm}
        \Statex \textbf{Input}: Resolvent operator $\mathcal{H}$, sparsity controlling parameter $\alpha$, accuracy $\epsilon$
        \vspace{0.14cm}
        \State \textbf{Initialization}: $\bsf^0$=random, with $\|\bsf^0\|_2=1$ and $\lambda^0$=$F(\bsf^k)$
        \vspace{0.05cm}
        \While {$|\lambda^{k+1}-\lambda^k|/\lambda^k > \epsilon$}
        \vspace{0.1cm}
        \State $\bsg_i^{k+1}=$sign$(\bmu_i^k)(\lambda^k|\bmu_i^k|-\alpha)_+$
        \vspace{0.1cm}
        \State $\bsf^{k+1}=\bsg^{k+1}/\|\mathcal{H}^* \bsg^{k+1} \|_2$
        \vspace{0.1cm}
        \State $\lambda^{k+1}=(1-\alpha)\|\bsf^{k+1}\|_2+\alpha \|\bsf^{k+1}\|_1$
        \vspace{0.1cm}
        \State $\bmu^{k+1}=\Sigma \bsf^{k+1}/\|\mathcal{H}^*\bsf^{k+1} \|_2$
        \vspace{0.1cm}
        \EndWhile
        \vspace{0.05cm}
        \Statex \textbf{Output}: raw sparse resolvent response modes $\bm \psi^{raw}=\bsf^{k+1}$
        \vspace{0.05cm}
\end{algorithmic}
\end{algorithm}

Lastly, as was indicated in \S\ref{sec:sparseResolvent}, for the cases presented in this manuscript, the authors have prescribed a set value of the sparsity ratio $\gamma$ instead of the sparsity ratio $\alpha$. In particular, the optimal value of $\alpha$ for a given value of the sparsity ratio $\gamma$ is determined by the process described in  algorithm~\ref{alg:alphaOptimal}, following \citet{hein2010inverse}. According to this, the sparse analysis identifies modes with a prescribed number of nonzero entries, for which the algorithm determines the optimal value of the sparsity parameter $\alpha$ that maximizes the sparsity above a pre-determined threshold $\beta$ while simultaneously maximizing the variance of the projection of a candidate eigenvector $\bsf$ onto $\mathcal{H}^*$. On the one hand, note that a larger $\alpha$ corresponds to more sparsity, and therefore a smaller number of nonzero terms (smaller $\gamma$). On the other hand, the variance of $\mathcal{H}^*\bsf$ is used as a measure of how well the relevant dynamics are captured with a chosen value of $\alpha$.

\begin{algorithm}
    \caption{Inverse power method for sparse resolvent analysis}\label{alg:alphaOptimal}
    \begin{algorithmic}[1]
        \vspace{0.1cm}
        \Statex \textbf{Input}: Resolvent operator $\mathcal{H}^*$, sparsity ratio $\gamma$, candidate eigenvector $\bsf$
        \vspace{0.14cm}
        \State \textbf{Initialization}: $\alpha^0$
        \vspace{0.05cm}
        \While{\{$\|w\|_1>\beta$ and Var$(\mathcal{H}^*\bsf)<\nu$\} }
        \vspace{0.1cm}
        \State \textbf{Optimize}: $w=\argmax \{$Var$(\mathcal{H}^*\bsw) -\alpha\|\bsf\|_1\}$
        \State \textbf{Evaluate}: is Var$(\mathcal{H}^*\bsf)>\nu$ and $\|\bsf\|_1<\beta$ ?
        \vspace{0.05cm}
        \State \textbf{Update}: Find next $\alpha$ using a grid search
        \vspace{0.1cm}
        \EndWhile
        \vspace{0.05cm}
        \Statex \textbf{Output}: optimal sparsity parameter $\alpha$
        \vspace{0.05cm} 
        \Statex Note that sparsity threshold $\beta$ and variance threshold $\nu$ are pre-set parameters
\end{algorithmic}
\end{algorithm}

\section{Convergence study of space-time resolvent analysis}\label{appB}
\begin{figure}
\centering {
\vspace{-0.cm}
{\hspace*{-1.0cm}\includegraphics[width= 1.14\textwidth]{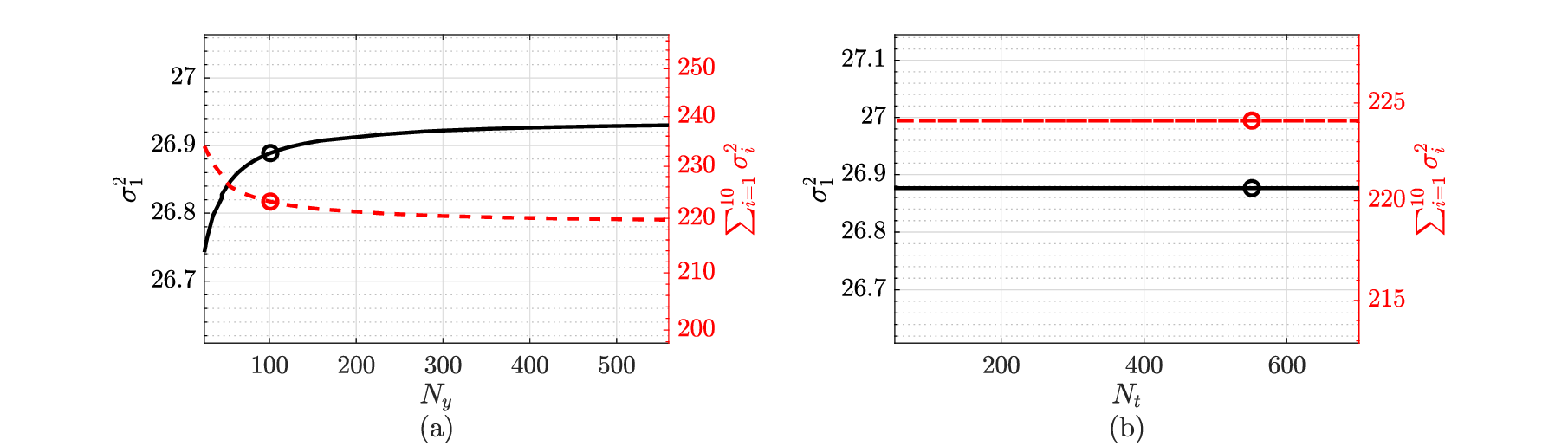}} }
\vspace{-0.3cm}
\caption{Evolution of $\sigma_1^2$ (black) and $\sum_i^{10} \sigma_i^2$ (red) as a function of (\textit{a}) $N_y$ for a constant total grid size $N=N_yN_t= 5.7\times 10^4$,  and (\textit{b}) as a function of $N_t$ for a constant spatial grid size $N_y=81$ in the non-sparse implementation of space-time resolvent analysis for a turbulent channel flow $\Rey_\tau=186$, and $\lambda_x^+=1000$, $\lambda_z^+=100$ for a time horizon $\tau_{tot}=20\tau=20(2\pi/\omega)$. Circles indicate resolution used in the rest of the paper.
}
\label{fig:convergence_turbchannel}
\vspace{-0.0cm}
\end{figure}

\begin{figure}
\centering {
\vspace{-0.cm}
{\hspace*{-1.0cm}\includegraphics[width= 1.14\textwidth]{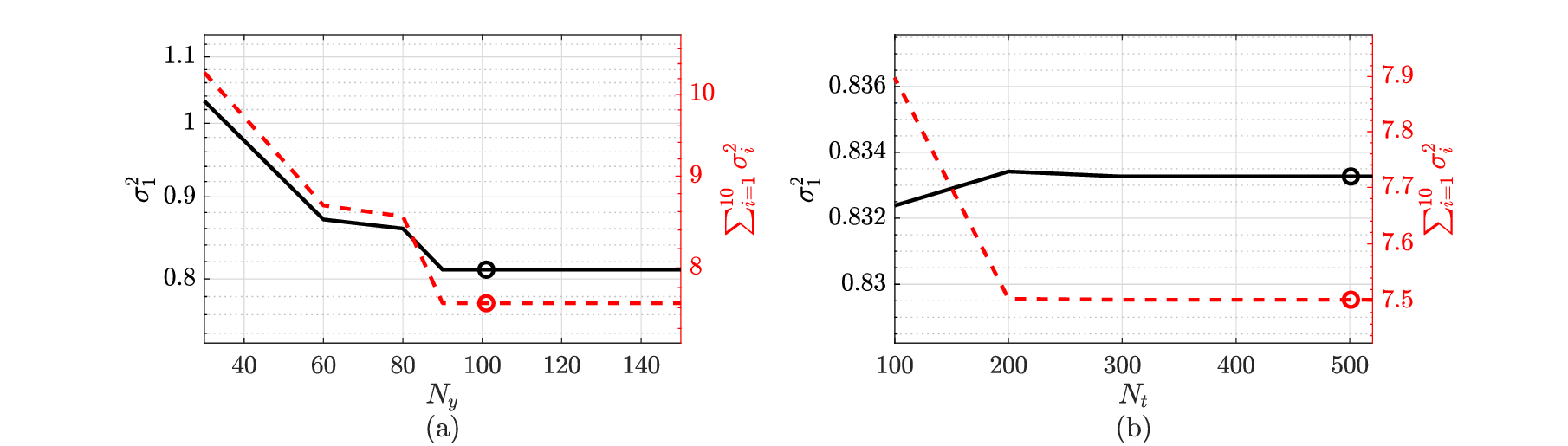}} }
\vspace{-0.3cm}
\caption{Evolution of $\sigma_1^2$ (black) and $\sum_i^{10} \sigma_i^2$ (red) as a function of $N_y$ for a constant total grid size $N=N_yN_t= 5.4\times 10^4$ (\textit{a}) and (\textit{b}) as a function of $N_t$ for a constant spatial grid size $N_y=101$, in the sparse implementation of space-time resolvent analysis for a turbulent channel flow $\Rey_\tau=186$, and $\lambda_x^+=1000$, $\lambda_z^+=100$ with $\gamma=0.001$ for a time horizon $\tau_{tot}=20\tau=20(2\pi/\omega)$. Circles indicate resolution used in the rest of the paper.
}
\label{fig:convergence_turbchannel_sparse}
\vspace{-0.0cm}
\end{figure}

\begin{figure}
\centering {
\vspace{-0.cm}
{\hspace*{-1.05cm}\includegraphics[width= 1.15\textwidth]{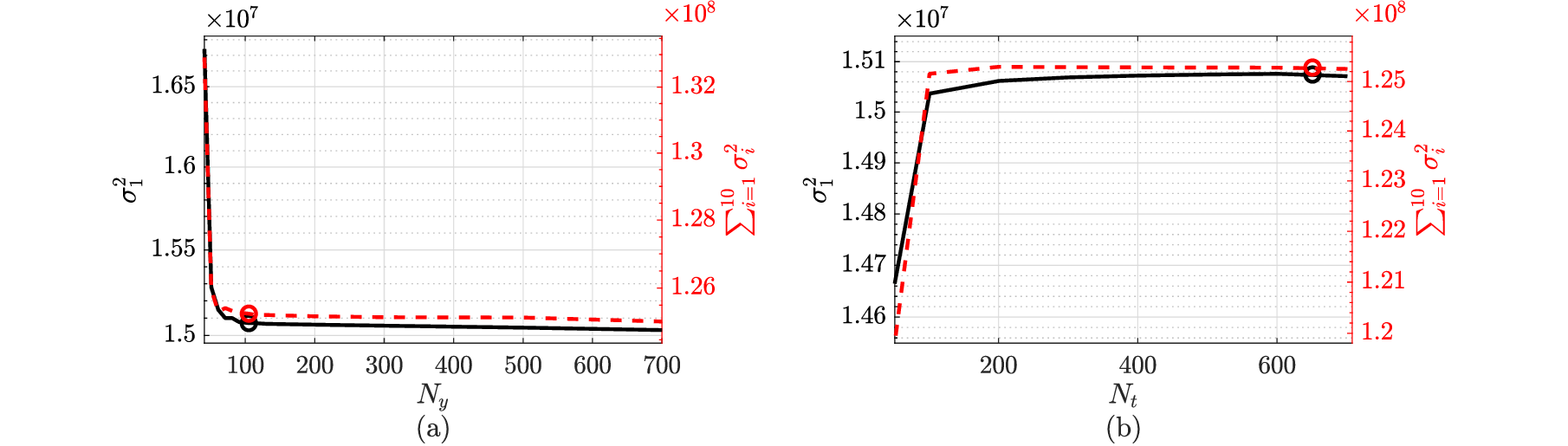}} }
\vspace{-0.3cm}
\caption{Evolution of $\sigma_1^2$ (black) and $\sum_i^{10} \sigma_i^2$ (red) as a function of (\textit{a}) $N_y$ for a constant total grid size $N=N_yN_t= 7.3\times 10^4$, and (\textit{b})  as a function of $N_t$ for a constant spatial grid size $N_y=81$, in the non-sparse implementation of space-time resolvent analysis for a turbulent Stokes boundary layer with $\lambda_x/h$=0.471 and $\lambda_z/h$=0.236 during a full oscillating cycle. Circles indicate resolution used in the rest of the paper.
}
\label{fig:convergence_stokes}
\vspace{-0.0cm}
\end{figure}

\begin{figure}
\centering {
\vspace{-0.cm}
{\hspace*{-1.05cm}\includegraphics[width= 1.15\textwidth]{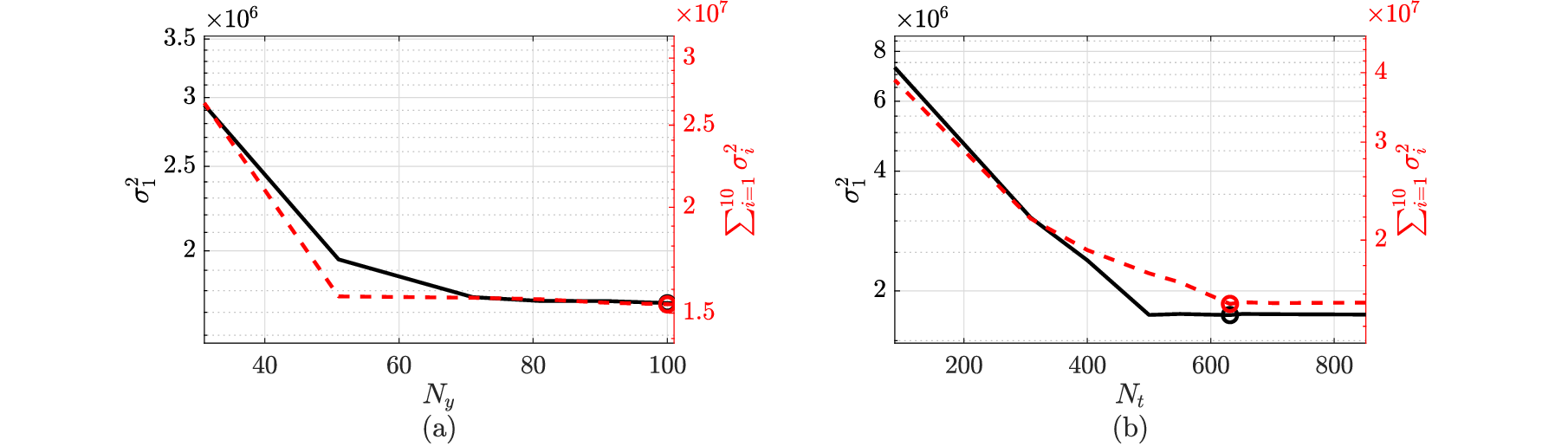}} }
\vspace{-0.3cm}
\caption{Evolution of $\sigma_1^2$ (black) and $\sum_i^{10} \sigma_i^2$ (red) as a function of (\textit{a}) $N_y$ for a constant total grid size $N=N_yN_t= 6.33\times 10^4$, and (\textit{b}) as a function of $N_t$ for a constant spatial grid size $N_y=81$, in the sparse implementation of space-time resolvent analysis for a turbulent Stokes boundary layer with $\lambda_x/h$=0.471 and $\lambda_z/h$=0.236 for $\gamma=0.001$ during a full oscillating cycle. Circles indicate resolution used in the rest of the paper.
}
\label{fig:convergence_stokes_sparse}
\vspace{-0.0cm}
\end{figure}

This appendix contains convergence studies related to  the results presented in this paper. In particular, a sweep was performed in a subset of the parameters $N_y$ and $N_t$. The convergence of the methods was assessed in terms of the energy contained by the leading mode $\sigma_1^2$, as well as the sum of the energy $\sum_i^{n} \sigma_i^2$ of the first $n=10$ modes. Figures \ref{fig:convergence_turbchannel}-\ref{fig:convergence_turbchannel_sparse} shows the convergence of the results for the statistically-stationary channel flow configuration, while figures \ref{fig:convergence_stokes}-\ref{fig:convergence_stokes_sparse} provides analogous results for the Stokes boundary layer. For the channel flow with a lateral pressure gradient, convergence was studied through comparison with results of \cite{ballouz2024wavelet}, as discussed in  \S\ref{sec:results3Dchannel}. 
Our available computational power (1,500 GB RAM) provided a threshold in terms of the total grid size $N=N_yN_t$ for each configuration. Note that the computational cost of the sparse formulation of the analysis is slightly higher, and therefore the maximum $N$ is slightly lower in these cases. In all cases, as indicated by circles on the plots, we choose the resolution $N_y$ and $N_t$ such that adding additional resolution in either dimension has minimal effect on both the leading and the sum of the squares of the first ten singular values.

\bibliographystyle{jfm}
\bibliography{Master}

\end{document}